\documentclass{aa}  
\usepackage{graphicx}
\usepackage{natbib}
\usepackage{amsmath}
\usepackage{wasysym}

\newcommand{\bm}{\boldsymbol}

\newcommand{\ob}{\overline}

\newcommand{\Rm}{R_{\rm m}}
\newcommand{\De}      {\mathrm{D}}
\renewcommand{\vec}[1]{\mbox{\boldmath{$#1$}}}

\newcommand{\Div}     {\vec{\nabla}\cdot}
\newcommand{\Laplace} {\nabla^2}

\newcommand{\Strain}{\mbox{\boldmath ${\sf S}$} {}}

\newcommand{\Av}      {\vec{A}}
\newcommand{\Bv}      {\vec{B}}

\newcommand{\cs}      {c_{\rm s}}

\newcommand{\aTens}{\mbox{\boldmath $\alpha$}}
\newcommand{\bTens}{\mbox{\boldmath $\beta$}}
\newcommand{\eTens}{\mbox{\boldmath $\eta$}}
\newcommand{\tkappa}{\chi_0}
\newcommand{\Rb}{\mathrm{Rb}}
\newcommand{\tAN}{t_{\rm A0}}
\newcommand{\vAN}{v_{\rm A0}}
\newcommand{\iunit}{\mathrm{i}}
\newcommand{\Tm}{{\mathrm{Ta}_\mathrm{M}}}
\newcommand{\sat}{\mathrm{sat}}
\newcommand{\figref}[1]{Fig.~\ref{#1}}
\newcommand{\betaZ}{\itilde{\beta}_0}
\newcommand{\crm}{\mathrm{c}}
\newcommand{\srm}{\mathrm{s}}
\newcommand{\kmax}{k_{\mathrm{max}}}

\newcommand{\EQ}{\begin{equation}}
\newcommand{\EN}{\end{equation}}
\newcommand{\EQA}{\begin{eqnarray}}
\newcommand{\ENA}{\end{eqnarray}}

\newcommand{\Eq}[1]{Equation~(\ref{#1})}

\newcommand{\Sec}[1]{Section~\ref{#1}}

\newcommand{\Fig}[1]{Figure~\ref{#1}}
\newcommand{\FFig}[1]{Figure~\ref{#1}}

\newcommand{\Tab}[1]{Table~\ref{#1}}

\newcommand{\bra}[1]{\langle #1\rangle}

{}
{}
{}
\newcommand{\meanemf}{\overline{\cal E} {}}

{}
{}
\newcommand{\meanEMF}{\overline{\mbox{\boldmath ${\cal E}$}}{}}{}
\newcommand{\meanemfs}{\overline{\cal E} {}}
\newcommand{\meanEEEE}{\overline{\mbox{\boldmath ${\cal E}$}}{}}{}
{}
{}
{}
{}
{}
\newcommand{\meanBB}{\overline{\mbox{\boldmath $B$}}{}}{}
{}
{}
{}
{}
{}
{}
{}
\newcommand{\meanJJ}{\overline{\mbox{\boldmath $J$}}{}}{}
\newcommand{\meanUU}{\overline{\mbox{\boldmath $U$}}}

{}
{}
{}

\newcommand{\meanB}{\overline{B}}

{}

{}
\newcommand{\emf}{{\cal E}}{}

\newcommand{\zz}{\mbox{\boldmath $z$} {}}

\newcommand{\kk}{\mbox{\boldmath $k$} {}}

\newcommand{\xx}{\mbox{\boldmath $x$}}

\newcommand{\uu}{\mbox{\boldmath $u$} {}}
\newcommand{\UU}{\mbox{\boldmath $U$} {}}

\newcommand{\bb}{\mbox{\boldmath $b$} {}}
\newcommand{\BB}{\mbox{\boldmath $B$} {}}

\newcommand{\JJ}{\mbox{\boldmath $J$} {}}

\newcommand{\nab}{\mbox{\boldmath $\nabla$} {}}
\newcommand{\OO}{\bm{\Omega}}

\newcommand{\dd}{{\rm d} {}}

\def\Ta{\mbox{\rm Ta}}

\def\Pm{\mbox{\rm Pr}_{\rm M}}
\def\Rm{\mbox{\rm Re}_{\rm M}}

\def\Rey{\mbox{\rm Re}}

\def\Lu{\mbox{\rm Lu}}
\def\csz{c_{\rm s0}}
\def\cs{c_{\rm s}}

\def\rms{{\mathrm{rms}}}

\newcommand{\yapj}[3]{ #1, {ApJ,} {#2}, #3}

\newcommand{\yapjl}[3]{ #1, {ApJ,} {#2}, #3}

\newcommand{\yan}[3]{ #1, {Astron.\ Nachr.,} {#2}, #3}

\newcommand{\yana}[3]{ #1, {A\&A,} {#2}, #3}

\newcommand{\yjfm}[3]{ #1, {J.\ Fluid Mech.,} {#2}, #3}

\newcommand{\yprs}[3]{ #1, {Proc.\ Roy.\ Soc.\ Lond.,} {#2}, #3}

\newcommand{\ybook}[3]{ #1, {#2} (#3)}
\newcommand{\yproc}[5]{ #1, in {#3}, ed.\ #4 (#5), #2}

\newcommand{\pjour}[2]{ #1, {#2}, to be published}

\newcommand{\itover}[2]{\,\hspace{.3mm}#1{\!\hspace{-.3mm}#2}}
\newcommand{\itilde}[1]{\itover{\widetilde}{#1}}
\newcommand{\ithat}[1]{\itover{\hat}{#1}}

\newcommand{\deriv}[3]{\frac{#3\hspace*{-.06em} {#1}}{#3\hspace*{.06em} {#2}}}

\newcommand{\parder}[2]{\deriv{#1}{#2}{\partial}}

\begin{document}
\title{Alpha effect due to buoyancy instability of a magnetic layer}
   \author{Piyali Chatterjee\inst{1}, Dhrubaditya Mitra\inst{1}, Matthias Rheinhardt \inst{1}, 
          \and
          Axel Brandenburg
          \inst{1,2}
          }

   \institute{NORDITA, AlbaNova University Center, Roslagstullsbacken 23,
              SE 10691 Stockholm, Sweden;
              \email{piyalic@nordita.org}
   \and
              Department of Astronomy, AlbaNova University Center,
              Stockholm University, SE 10691 Stockholm, Sweden
             }
   \titlerunning{Buoyancy instability of magnetic layer}
   \authorrunning{P. Chatterjee et al.}

   \date{$ $Revision: 1.237 $ $}

 
  \abstract
   {
    A strong toroidal field can exist in form of a magnetic layer in the overshoot 
    region below the solar convection zone.
    This motivates a more detailed study of the magnetic buoyancy
    instability with rotation.
   }
   {
    We calculate the $\alpha$ effect due to helical motions caused by a
    disintegrating magnetic layer in a rotating density-stratified system
    with angular velocity
    $\Omega$ making an angle $\theta$ with the vertical. We also study the     
    dependence of the $\alpha$ effect on $\theta$ and the strength of 
    the initial magnetic field. 
    }
   {
    We carry out three-dimensional hydromagnetic simulations in
    Cartesian geometry.
    A turbulent EMF due to the correlations of the 
    small scale velocity and magnetic field is generated. We use the test-field
    method to calculate the transport coefficients of the inhomogeneous turbulence
    produced by the layer.
    }
   { We show that the growth rate of the instability and the twist of the magnetic 
   field vary monotonically with the ratio of thermal conductivity to magnetic 
   diffusivity. 
     The resulting $\alpha$ effect is inhomogeneous
     and increases with
     the strength of the initial magnetic field.   
     It is thus an example of an ``anti-quenched'' $\alpha$ effect.
     The $\alpha$ effect is nonlocal, requiring around 8--16 Fourier
     modes to reconstruct the actual EMF based on the actual mean field.
   }
   {}
   \keywords{magnetohydrodynamics (MHD) -- Sun: magnetic fields --
   Instabilities -- Turbulence -- Sun: dynamo}

\maketitle
%

\section{Introduction}

The magnetic fields in many astrophysical bodies have their origin
in some kind of turbulent dynamo.
This means that
a part of the kinetic energy of the turbulent motions
is diverted to enhancing and maintaining a magnetic field.
This magnetic field is generally also random, but under certain
conditions a large-scale magnetic field can also emerge.
Here by large-scale we mean length scales larger than the
energy containing scale of the fluid. 
This can be the case when the turbulence is helical, 
e.g., owing to the simultaneous presence of rotation and stratification.

The evolution of the large-scale magnetic field can be described
using averaged evolution equations.
In the process of averaging, new terms emerge (e.g., the $\alpha$ effect
and turbulent diffusion) that result from correlations
between small-scale velocity and magnetic fields.
Here one usually considers the case where the magnetic fluctuations
are caused by the fluctuating velocity acting on the mean field.
However, under certain conditions it might well be the other way around.
Imagine, for example, the case where initially no velocity is present, but 
there is instead a strong large-scale magnetic field the presence of which
makes the initial state of zero velocity unstable. 
In that case the magnetic field would be responsible for driving
velocity and magnetic fluctuations at the same time.
This type of scenario was first simulated in the context of accretion discs
where the magneto-rotational instability drives turbulence \citep{BNST},
and later in the context of the magnetic buoyancy instability with shear
\citep{CBC03}, which might apply to the overshoot layer of the Sun.
It had already been proposed by \cite{Mof78} that,
once the dynamo-generated magnetic field
in this layer reaches appreciable strengths,
the magnetic buoyancy
instability can set in and govern the dynamics thereafter.
The linear phase of this instability in a 
localized flux layer with stratification
and rotation was later studied in detail by \cite{Sch84,Sch85}. 
A necessary but not sufficient condition for
this instability is
\begin{equation}
\frac{\partial}{\partial z}\log\left(\frac{B}{\rho}\right) < 0,
\label{InstCrit}
\end{equation}
which essentially means that the magnetic field
modulus $B$ decreases faster
with height $z$ than the density $\rho$.
\cite{BS98} performed numerical calculations
in presence of rotation
and determined the $\alpha$ effect of the resulting
turbulence by imposing an external magnetic field.

This type of magnetic buoyancy instability
is also related to the undulatory instability in the 
absence of both rotation and shear \citep{Fan01}
and the double diffusive instability \citep{Silvers} 
in presence of shear and no rotation.
While the focus of the first
study has been on the formation of flux tubes 
from a pre-existing toroidal magnetic layer in a stably stratified atmosphere, 
in the latter a magnetic layer was generated from an 
initially vertical magnetic field in 
presence of strong shear.
It was further shown that, when the ratio of 
magnetic to thermal diffusivities is sufficiently low, magnetic 
buoyancy can still operate in the tachocline.

The focus of this work is twofold.
Firstly, we want to study the nature of the instability
at short times, i.e., in its initial linear stage. 
In particular, its
dependence on various parameters such as magnetic and thermal
Prandtl numbers, angular velocity, strength of the 
initial field, etc, and compare against the linear theory and 
previous numerical work.
It can be argued that in presence of rotation this instability produces 
magnetostrophic waves due to balance between Coriolis 
and Lorentz forces. 
An important result highlighted later is that rotation is 
not vital to the growth of this instability.
Secondly, we want to study whether this instability constitutes
a viable dynamo process, so we want to
measure the mean-field transport coefficients, namely the tensors 
$\aTens$ and $\eTens$ using the quasi-kinematic
test-field (QKTF) method \citep{Sch05,Sch07}.
However, with one exception \citep{VB09},
the QKTF has never been applied to the 
calculation of transport coefficients 
in an inhomogeneous turbulence induced by the  
mean magnetic field itself. Therefore we aim to first
verify the applicability of the QKTF method to this problem. For a review 
on transport coefficients and their determination using test fields;
see \cite{Betal10}.
The applicability of this  method to problems with an initial magnetic
field and fluctuations generated from it is discussed in \cite{RB10}.
\section{The Model} 
\begin{figure}
\includegraphics[width=0.5\textwidth]{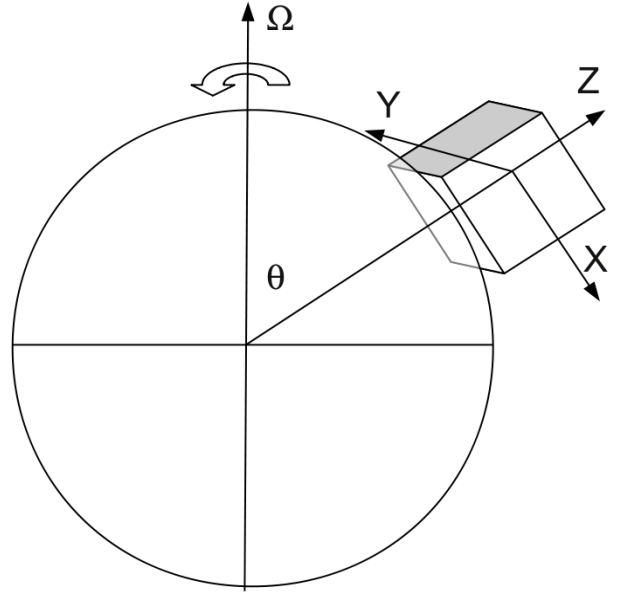}
\caption{\label{fig:cartoon} The Cartesian simulation domain with respect to
spherical coordinates.}
\end{figure}

We consider a setup similar to that described in 
\cite{BS98}. 
The computational domain is a cuboid
with constant gravity, $g_z$, pointing in the negative $z$ direction,
and rotation $\OO$ making an angle $\theta$ with the vertical. 
The box may be thought to be placed at a colatitude $\theta$ on the 
surface of a sphere with its unit vectors
$\bm{\hat{x}}, \bm{\hat{y}}, \bm{\hat{z}}$ pointing along the 
local $\theta, \,\phi,\, r$ directions, respectively,
as shown in Fig.~\ref{fig:cartoon}.

We solve the following set of MHD equations.
The continuity equation is given by
\begin{equation}
\label{eq:continuity}
\frac{\De\ln\rho}{\De t}
  = - \Div\UU,
  \end{equation}
where
$\De/\De t \equiv \partial/\partial t + \UU\cdot\nab$ denotes the Lagrangian derivative
with respect to the local velocity of the gas $\UU$.
Assuming an ideal gas, we express the pressure in terms of density,  
specific entropy $s$, and sound speed $\cs$, which, in turn,
is a function of $\rho$ and $s$.
Thus the momentum equation
in a frame of reference rotating with angular velocity $\OO$ reads
\EQ
  \begin{aligned}
    \frac{\De\UU}{\De t}
   =&-\cs^2\nab\biggl(\frac{s}{c_p} + \ln\rho\biggr)
   - 2 \OO\times \UU   
      + g_z \vec{\hat{z}}                         
      + \frac{\JJ\times\Bv}{\rho}  \\
      &+ \nu \left( \Laplace\UU + \frac{1}{3}\nab\Div\UU
      + 2\Strain\cdot\nab\ln\rho\right),
\end{aligned} \label{eq:NS}
\EN
where $\JJ$ is the current density, $\Bv$ is
the magnetic field, $\nu$ is the constant kinematic viscosity,
and $\Strain$ is the traceless rate-of-strain tensor.
The sound speed is related to temperature by $\cs^2 = (c_p-c_v)\gamma T$
with $c_p$ and $c_v$ the
specific heat at constant pressure  
and constant volume, respectively, and $\gamma=c_p/c_v$ is here fixed to $5/3$.
The induction
equation is solved in terms of the magnetic vector potential $\Av$, 
such that
$\nab\times\Av = \Bv$, hence
\begin{equation}
  \frac{\partial\Av}{\partial t}
  = \UU\times\Bv + \eta\nabla^2 \Av \,, \\
\end{equation}
where $\eta$ denotes constant molecular magnetic
diffusivity. 

Finally, we have for the entropy equation with 
temperature $T$ and constant radiative (thermal) conductivity $K$
\begin{equation}
\label{eq:entropy}
   \rho T\frac{\De s}{\De t}
   =  \Div(K\nab T)
      + \eta\mu_0\JJ^2
      + 2\rho\nu\Strain^2 \,,
\end{equation}
where the temperature is related to the specific entropy by
\EQ
s = s_0 + c_v \ln\frac{T/T_0}{(\rho/\rho_0)^{\gamma-1}}\,.
\label{entropy}
\EN
We use the fully compressible {\sc Pencil Code}\footnote{
http://www.pencil-code.googlecode.com} for all our calculations. 

For all quantities, periodic boundary conditions in the $x$ and $y$ directions are 
adopted. In the $z$ direction we use the no-slip
boundary condition for the velocity,
the vertical field condition for the magnetic field, as a proxy for vacuum boundaries.
We keep the temperature at the top and the (radiative) heat
flux at the bottom fixed.        
Their values were chosen to conform with the initial temperature profile
of the (not magnetically  modified) polytrope  described below.

\subsection{Initial state}

The base state is a polytrope 
that is, $p=C\rho^\Gamma$, with index $m=1/(\Gamma-1)=3$.
The initial $z$ profiles of density, pressure, temperature and entropy are given by,
\EQ
\begin{aligned}
\rho_i &= \rho_0\Phi^3(z), \quad 
p_i      = p_0\Phi^4(z), \quad 
T _i     = T_0\Phi(z), \\
s_i & = s_0 -  c_v\ln \Phi(z), 
\end{aligned}\label{polytr}
\EN
where $\Phi$ is a non-dimensional gravitational potential given by
\EQ
\Phi(z)  = 1+\frac{1}{4}\frac{g_z}{T_0(c_p-c_v)} (z-z_0),
\nonumber
\EN
with the reference point $z_0$ chosen to be at the bottom of the domain
and the values at this point given by $\rho_0$, 
$p_0=c_{s0}^2\rho_0/\gamma$, $T_0=c_{s0}^2/(c_p-c_v)\gamma$ and
 $s_0$. Here $\csz$ is the reference sound speed to which we also refer to
when calculating Mach numbers.

As the adiabatic index is here $m_{\rm ad}=1/(\gamma-1)=3/2$,
the subadiabaticity in the domain is very large, namely
$\partial \ln T/\partial \ln P -( \partial \ln T/\partial \ln P)_{\rm ad}=-0.15$.
Thus, the initial stratification is
highly stable to convection in the absence of any magnetic field,
guaranteeing that turbulence is generated solely 
by the buoyancy instability.

The initial magnetic field is a horizontal
layer of thickness $H_B$, where $B_y$ has the profile
\begin{equation}
B_{y0} = B_0 H_{\rm B}\parder{}{z}\tanh\left(\frac{z-z_B}{H_{\rm B}}\right),
\end{equation}
and the reference Alfv\'en speed is defined by
$v_{\rm A0} = B_0/ \sqrt{\rho_0 \mu_0}$
with $\mu_0$ being the vacuum permeability.
If not indicated otherwise,  the initial magnetic field strength is fixed to
$v_{\rm A0}/c_{\rm s0}=0.5$.
In order to satisfy the condition \eqref{InstCrit} initially, we have to ensure
$H_B<H_\rho(z_B)$, where $H_\rho(z)=|\nab\ln\rho(z)|^{-1}$ is the local density
scale height.
When choosing $z_B-z_0=0.3 L_z$
this is satisfied for 
$H_B < 0.1L_z + 4 T_0 (c_p -c_v)/3 |g_z|$
which is surely true for the choice $H_B=0.05 L_z$.

Upon addition of a magnetic field, we
have to modify the base state such that the 
density profile remains unchanged.
In order to obey magnetostatic equilibrium,
pressure and temperature are adjusted in the following way:
\EQ
p_i \Rightarrow p_i - \frac{B_{y0}^2}{2\mu_0}, \qquad T_i \Rightarrow T_i - \frac{B_{y0}^2}{2\mu_0} \frac{1}{\rho_i(c_p-c_v)}. \label{polytrmod}
\EN
The entropy is then re-calculated from Eq.~(\ref{entropy}).
The initial velocity components $U_x$ and $U_y$
are specified such that it contains about 20 localized eddies
in the plane $z=z_B$
with Mach  
numbers of about $10^{-5}$. Also the initial vertical velocity, $U_z$ is 
Gaussian random noise with the same Mach number.
The rms of the initial kinetic helicity,
scaled with the product of initial rms velocity and vorticity,
is denoted $\varepsilon_{\rm K0}$,
that is, $\varepsilon_{\rm K0} = (\sqrt{\bra{(\bm W\cdot \bm U)^2}} / U_\rms W_\rms)(0) = 4\times 10^{-6}$.

\subsection{Control parameters, nondimensional quantities,
and computational grid}
\label{params}
The problem posed by \eqref{eq:continuity} through \eqref{eq:entropy} is
governed by five independent dimensionless parameters, (i) the Prandtl number
$\Pr=\nu/\tkappa$, with the temperature conductivity $\tkappa=K/\rho_0 c_p$,
(ii) the magnetic Prandtl number $\Pm=\nu/\eta$,
(iii) the ``magnetic Taylor number" $\Tm = 2 \Omega^2 L_y^4/\eta^2$, 
(iv) the rotational inclination (colatitude),
$\theta$,
and (v) the normalized gravitational acceleration $g_z L_y^3 \eta^2$.
In addition there are two independent parameters of the initial equilibrium
(vi) the normalized pressure scale height at the bottom, $H_P/ L_z=c_{s0}^2/\gamma g_z L_z$
and (vii) the initial Lundquist number, $\Lu_0=\vAN H_B / \eta$,
based upon the thickness of the magnetic layer.
In addition to this, we also have the non-dimensional sound speed, $c_{s0}L_y/\eta$.
In this paper we shall keep the normalized pressure scale height and 
the sound speed fixed, 
while varying 
both Prandtl numbers, $\Tm$, $\theta$ and $\Lu_0$.
The definitions as well as the values or ranges of the control parameters are 
summarized in \Tab{tab:parms}.
We have also included in the same table two dependent parameters namely
the modified initial plasma-beta in the midplane of the magnetic layer
and the
{\em Roberts number}
$\Rb=\Pm/\Pr = \tkappa/\eta$.

The computational domain is defined by
$|x| \le L_x/2$, $|y|\le L_y/2$,  $-L_z/4\le z\le 3 L_z/4$, $L_x=L_z=L_y/3$,
thus its aspect ratio is 1:3:1.
The results will be presented in non-dimensional form, velocity in units
of the reference Alfv\'en speed, $v_{\mathrm A0}$,
time in units of the corresponding Alfv\'en travel time in 
the $y$ direction, $\tAN= L_y/v_{\mathrm A0}$, and
magnetic field in units of $B_0$ or the rms value
$(\int_z B_{y0}^2 \dd z /L_z)^{1/2}$.

It is instructive to look upon the relevant definitions of the fluid Reynolds
number, $\Rey$, and the magnetic Reynolds number, $\Rm$, for this problem where
the turbulence is driven solely by the instability of the magnetic layer.
From first principles, the $\Rey$ characterizes the
ratio of the advective term $\bra{\bm{(U\cdot\nabla U})^2}^{1/2}$
and the viscous term
$\bra{(\nu\nabla^2\bm U)^2}^{1/2}$ in the Navier-Stokes equation, 
while $\Rm$ characterizes the ratio of 
$\bra{\big(\bm{\nabla}\times(\bm U\times\bm B)\big)^2}^{1/2}$ and  
$\bra{(\eta\nabla^2\bm B)^2}^{1/2}$ in the induction equation with the 
angular brackets representing
volume averaging.
Let us denote these ab initio definitions as ``term-based" and refer to them by 
$\Rey^*$ and $\Rm^*$.
Note, that with the term-based definitions $\Rm/\Rey$ may well deviate from $\Pm$.
Alternatively, we can define a length scale 
$L_U = U_{\rm rms}/2\pi W_{\rm rms}$
from the rms values of velocity and vorticity and define
the more conventional ``length-based'' Reynolds numbers 
$\Rey=U_\rms L_U/\nu$ and $\Rm = U_\rms L_U/\eta$. 

The calculations were carried out 
on equidistant grids with
resolutions of either $64^3$ or $128^3$. For numerical testing we have also
performed a few runs with $256^3$ or $128^2\times256$ resolutions.
\begin{table}
\caption{\label{tab:parms}
Non-dimensional control parameters
characterizing the buoyancy instability. Note the definition of the modified plasma-beta $\itilde{\beta}$ as the ratio of the total pressure $p_{\mathrm{tot}}=p + p_{\mathrm{M}}$
to the magnetic pressure $p_{\mathrm{M}}= B_{y0}^2/2\mu_0$, because this quantity adopts a simple $1/B^2$ dependence on the magnetic field, cf Eq.~\eqref{polytrmod}.
Values of $\itilde{\beta}$
refer to $t=0$ and the midplane of the magnetic sheet.}
\begin{tabular}{@{\hspace{0mm}}l@{\hspace{4mm}}l@{\hspace{.6mm}}c@{\hspace{.5mm}}c@{\hspace{0mm}}}
\hline
\hline
\\[-1mm]
Parameter & \hspace*{-3mm}Symbol & Definition& Value/Range\\[1mm]
\hline
\\[-2mm]
norm. scale height &  & $H_P/L_z$& 0.3\\[1mm]
norm. sound speed & & $c_{s0}L_y/\eta$ & $6\times 10^4$\\
Prandtl number & $\rm Pr$ & $\nu/\tkappa$&0.125 \ldots 4.0\\[1mm]
magnetic Prandtl no.\!\! & $\rm Pr_{\rm M}$ &$\nu/\eta$& 0.125 \ldots 4.0\\[1mm]
Roberts number & $\rm Rb$ &$\tkappa/\eta$& 0.25 \ldots 1.0\\[1mm]
magnetic Taylor no. & $\Tm$ & $\Omega^2L_{\rm y}^4/\eta^2$ & $0 \ldots 3.2\times10^{10}$\\[1mm]
rotational inclination & $\theta$ & $\varangle(\OO,\ithat{\zz})$  & 0 \ldots 180\\[1mm]
(initial)
Lundquist no. & $\rm Lu_0$ & $v_{\rm A0}H_{\rm B}/\eta$& 500 \ldots 600\\[1mm]
(initial) modified & $\betaZ$ & $(p_{\mathrm{tot}}/p_{\mathrm{M}})(z_{\rm B},0)$&1.04 \ldots 3.22 \\[-0.5mm]
plasma-beta\\[1mm]
\hline
\end{tabular}
\end{table}
\subsection{The test-field method}

\begin{table*}[t]
\caption{\label{tab:res1}List of runs of set B.
The computational box is placed at colatitude $\theta=30^\circ$.
Magnetic Taylor number $\Tm=3.24\times10^{10}$,
initial Lundquist number $\Lu_0=500$, initial plasma-beta $\betaZ=2.27$ 
and resolution $128^3$ throughout.
$\rm Ma$ -- Mach number, based on $U_\rms$,  
$\omega_\mathrm{I}$ -- growth rate.
Saturation reached at $t^{\rm sat}$.
For the mean EMF in the saturated stage
global extrema of the dominating $\emf^{\rm sat}_y$  with respect to $z$ and $t$ are given.
}
\centering
\begin{tabular}{@{}l@{\hspace{4mm}}   c @{\hspace{3.5mm}} c c @{\hspace{3mm}}c @{\hspace{3.5mm}}c @{\hspace{5.5mm}} c c@{\hspace{10.5mm}} c c @{\hspace{6.5mm}}c  c@{\hspace{3.5mm}} c@{\hspace{-2mm}}}     
\hline\hline\\[-2mm]
\phantom{}Run &  $\Pr$ & $\Pm$ & $\rm Ma$ & \hspace{0mm}$\omega_\mathrm{I}\,\tAN$ & $t^{\rm sat}/\tAN$ & 
\multicolumn{2}{c}{\hspace*{-7mm}$10^4 \times \emf_y^{\rm sat}/\vAN B_0$} & \multicolumn{1}{c}{\hspace*{-1mm}length-based}  &  \multicolumn{3}{c}{term-based}\\   
&&&&&&{\hspace*{0mm}min} & {\hspace*{0mm}max}& $\Rey\,(L_U)$ & {\hspace*{1mm}$\Rey^*$} & {\hspace*{-1mm}$\Rm^*$} & {\hspace*{-0mm}$\Rm^*/\Rey^*$}\\[1mm]  
\hline\\[-2mm]
   B128a &  4.0     & 4.0     & 0.017    & 15.6   & 1.99  &$-1.01$&2.34&0.5&0.4&2.3&5.8\\	 
   B128b &  1.0     & 4.0     & 0.036   & 21.6   & 1.42  &$-3.39$& 7.32 &0.9&0.6&2.8& 4.5\\	 
   B128c &  1.0      & 1.0    & 0.020  & 13.2  & 1.64 &$-1.49$&3.03 &1.8&1.4&1.9&1.4\\	 
   B128d &  0.25   & 1.0      & 0.038   &  25.2     & 1.27  &$-4.02$&7.52 &2.9&2.1&2.8&1.3\\	 
   B128e &  0.125 & 0.5       & 0.036 & 24.0  & 1.22  &$-5.47$&6.19& 3.6&3.3&2.9&0.9\\	 
   B128f  &   0.125 & 0.125   & 0.043   & 19.9  & 1.54  &$-3.50$&4.84 & 8.2&16.1&3.1&0.2\\	 
   B128g &  0.5     & 0.5     & 0.018   &  19.2 & 1.72  &$-2.06$&3.69 & 2.9&2.5&1.9&0.8\\	 
   B128h &  0.5     &1.0      &0.032 &  21.6 &1.67   &$-3.94$& 3.97& 1.7&1.9&3.2&1.7\\[1mm]	 
\hline
\end{tabular}
\end{table*}

We now define mean magnetic and velocity fields, $\meanBB$ and $\meanUU$,
where overbars denote horizontal averaging.
Fluctuations are defined correspondingly as $\bb=\BB-\meanBB$ and
$\uu=\UU-\meanUU$.
Following the above
convention, the induction equation may be horizontally averaged as,
\begin{equation}
\parder{\meanBB}{t}=\nab\times\left(\meanUU\times\meanBB\right)
+\nab\times\meanEMF +\eta\nabla^2\meanBB,
\label{eq:dBdt}
\end{equation}
where $\eta$ is the molecular magnetic diffusivity of the fluid
(here assumed uniform), while $\meanEMF\equiv\overline{\uu\times\bb}$
is the mean electromotive force. The essence of mean-field magneto-hydrodynamics is to 
provide an expression for $\meanEMF$ as a function of the large scale magnetic field 
and its derivatives. Mathematically, 
\begin{equation}
\meanEMF=\aTens\meanBB - \eTens\nab \meanBB, \label{eq:Eansatz}
\end{equation}
where $\aTens$ and $\eTens$ are called transport coefficients.
Note that a much more general representation of $\meanEMF$ is given 
by the convolution integral
\EQ
   \meanEMF(\xx,t) = \int_{t_0}^t \int \boldsymbol{G}(\xx,\xx',t,t')\, \meanBB(\xx',t') \, \dd^3 x' \, \dd t' \label{convol}
\EN
with an appropriate tensorial kernel $\boldsymbol{G}$. The aim of the 
test-field method is to provide an expression for $\boldsymbol{G}$ as a function 
of fluid properties.
By subtracting the horizontally averaged equation 
from the real one, we obtain the following
equation for the fluctuating magnetic field $\vec{b}$.
\begin{equation}
\label{eq:dbdt}
{\partial\bb^{pq}\over\partial t}=\nab\times\left(\meanUU\times\bb^{pq}
+\uu\times\meanBB^{pq}+\vec{e}^{pq}\right) +\eta\nabla^2\bb^{pq},
\end{equation}
with,  $\bm{e}^{pq}=\uu\times\bb^{pq} -\meanEMF^{pq}$.
The superscripts $pq$ indicate that this equation is solved
for suitably chosen test fields $\meanBB^{pq}$ with $p, q=1, 2$ if $\aTens$ 
and $\eTens$ are assumed to be $2\times 2$ matrices. This is the equation 
invoked by the test-field method
for calculating the tensors $\aTens$ and $\eTens$. 
The test-field suite of the {\sc Pencil Code} has the provision for using either
harmonic test fields i.e.,
\begin{align}
&\begin{aligned}
\meanB^{11}&=(\cos kz, 0, 0),\quad \meanB^{12}=(0, \cos kz, 0),\\ 
\meanB^{21}&=(\sin kz, 0, 0),\quad \meanB^{22}=(0, \sin kz, 0),
\end{aligned}\\
\intertext{or linear test fields i.e., }
&\begin{aligned}
\meanB^{11}&=(1, 0, 0),\quad \meanB^{12}=(0, 1, 0),\\
\meanB^{21}&=(z, 0, 0),\quad \meanB^{22}=(0, z, 0).
\end{aligned}
\end{align}

When it comes to applying the test-field method, an aspect not discussed
up to now is the intrinsic 
inhomogeneity of the flow
both due to stratification and the background magnetic field itself.
Within kinematics, that is without the background field,
no specific complication is connected to this as
$\aTens$ and $\eTens$ emerge straightforwardly
from the stationary version of \Eq{convol}
in a shape expressing inhomogeneity, that is, 
$\aTens(\xx,\xx')$, $\eTens(\xx,\xx')$
or, equivalently, $\aTens(\xx,\xx-\xx')$, $\eTens(\xx,\xx-\xx')$.
Performing a Fourier transform with respect to their second argument,
we arrive at $\hat{\aTens}(\xx,\kk)$ and $\hat{\eTens}(\xx,\kk)$.
In our case, harmonic test fields with different wavenumbers $k$
in the $z$ direction can be employed to obtain $\hat{\aTens}(z,k)$
and $\hat{\eTens}(z,k)$.

In the nonlinear situation,
the Green's function approach remains valid if $\meanEMF$ is considered
as a functional of $\UU$ and $\meanBB$ which is then linear and 
homogeneous in the latter.
However, we have to label $\boldsymbol{G}$ by the $\meanBB$
actually acting upon $\UU$,
that is, $\boldsymbol{G}(\xx,\xx';\meanBB)$, and can thus only make statements
about the transport tensors for just the particular $\meanBB$ at hand.
Hence, the tensors have to be labelled likewise: $\hat{\aTens}(z,k; \meanBB)$,
$\hat{\eTens}(z,k; \meanBB)$.
As our initial mean magnetic field is in the $y$ direction,
the instability will generate a $\meanB_x$
and we are mainly interested in
the coefficients
$\alpha_{21}$, $\alpha_{22}$, $\eta_{21}$ and $\eta_{22}$
with rank-2 tensor components $\eta_{ij}=-\eta_{ik3}\epsilon_{jk3}$.

\section{Results}

\subsection{Nature of the instability}

To start with we have performed a number of runs with different values
of $\Pr$ and  $\Pm$, but all other dimensionless parameters held fixed,
see \Tab{tab:res1}.
In particular, we have used a value of  $\Tm=3.24\times10^{10}$
for the magnetic Taylor number
and $\Lu_0=500$ for the initial Lundquist number.
\Tab{tab:res1} shows the Reynolds numbers according to the two alternative definitions 
provided in \Sec{params}. Note that with the exception of the run B128f, $\Rey$ from 
the ``length-based'' and the ``term-based'' definitions are in agreement. 
Also the ratio $\Rm^*/\Rey^*$ from the term-based definitions approaches 
$\Pm$ reasonably.

\begin{figure}[t]
\includegraphics[width=0.5\textwidth]{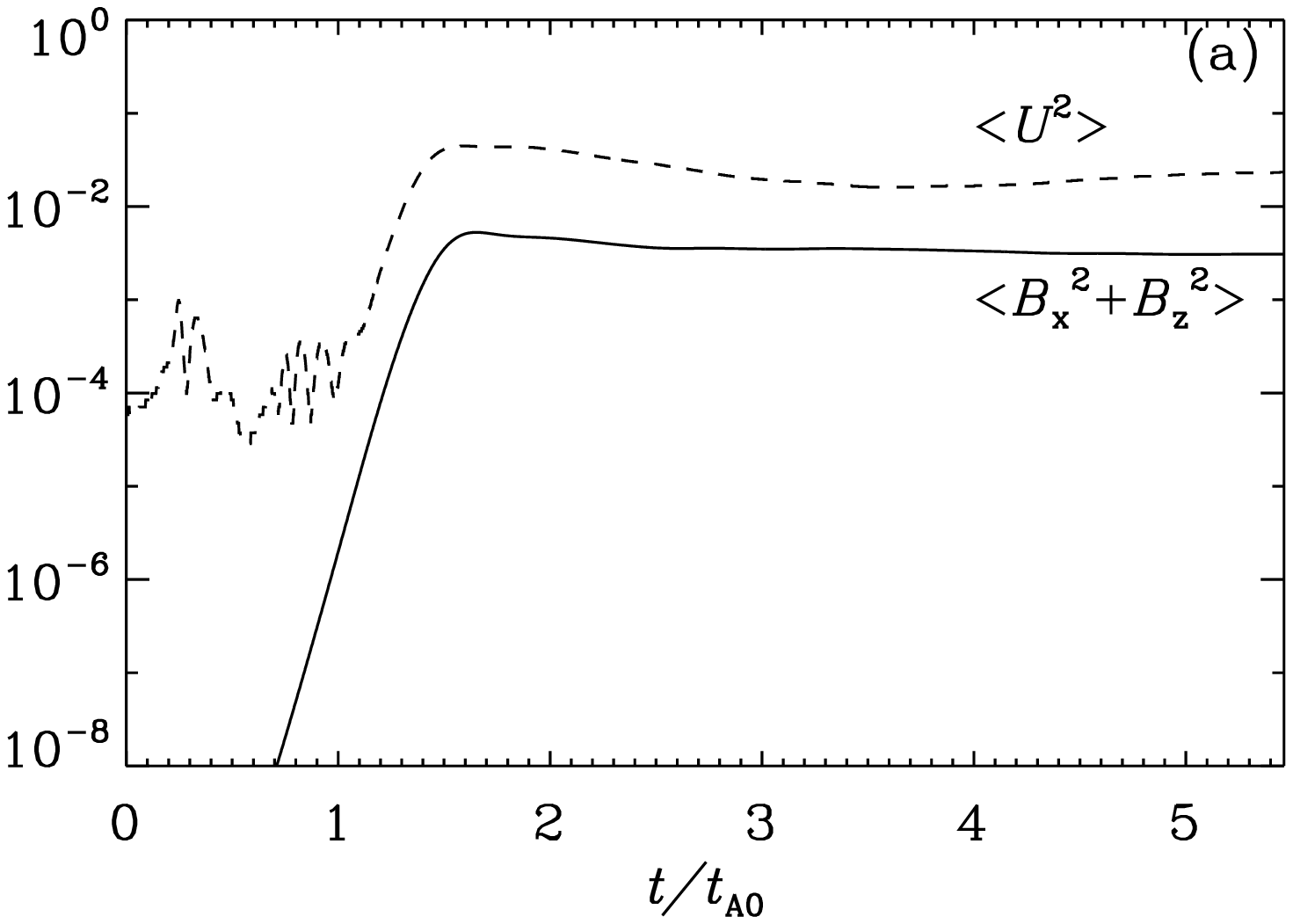}  
\includegraphics[width=0.5\textwidth]{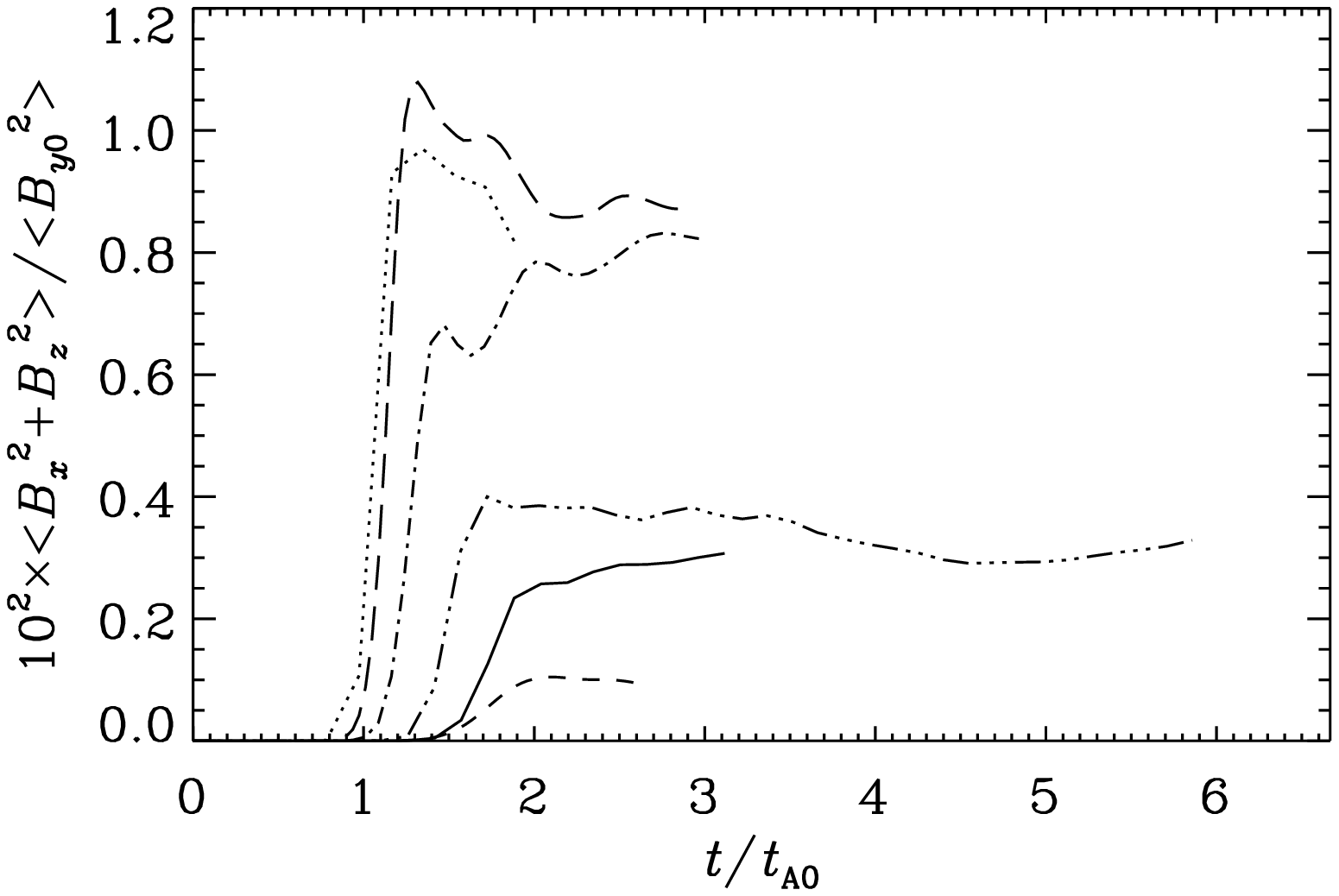}
\includegraphics[width=0.5\textwidth]{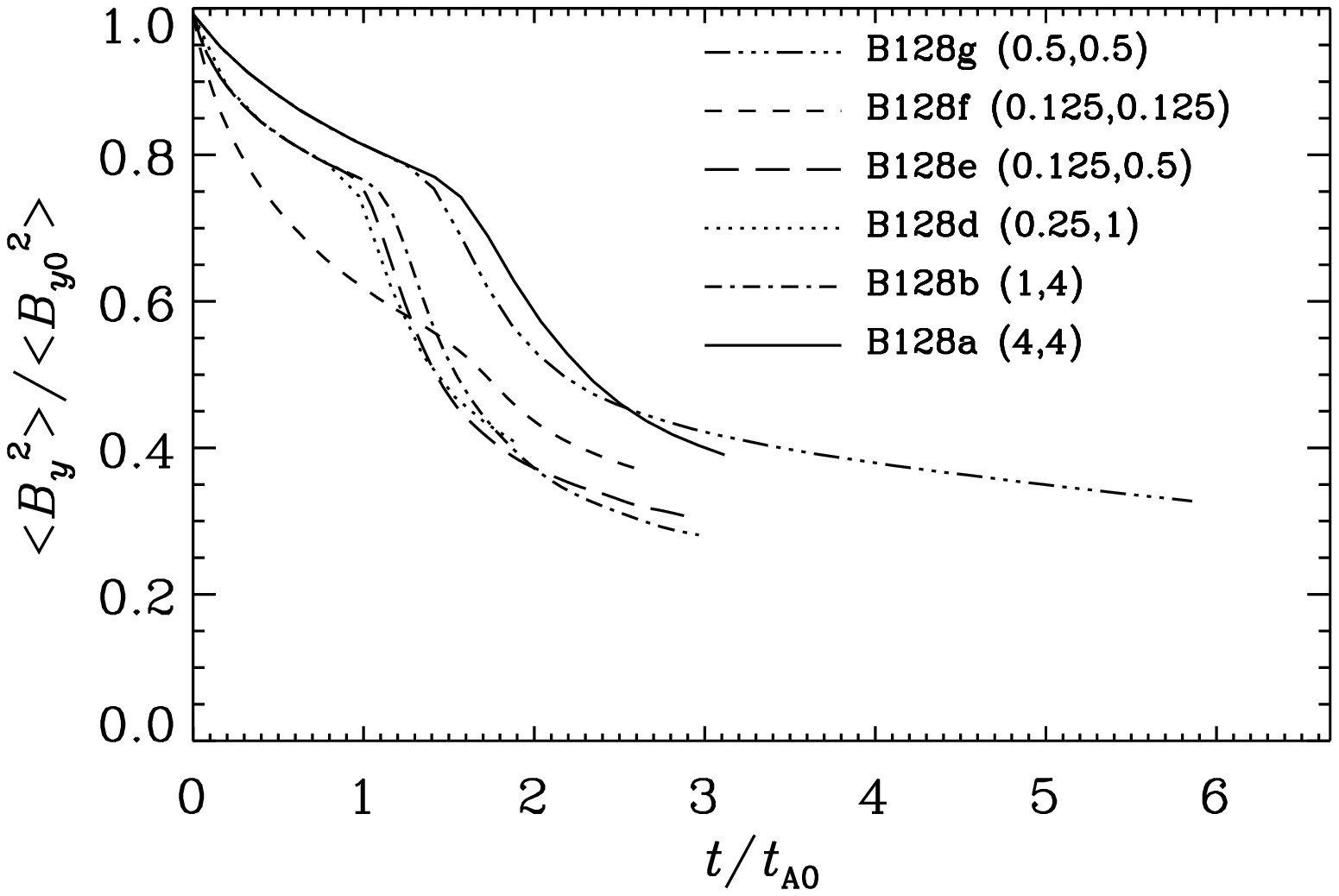}
\caption{\label{fig:energetics}
Time evolution of the runs in \Tab{tab:res1}.
{\em Upper panel:} rms values of velocity and generated magnetic field components $B_x$, $B_z$ of run B128c \mbox{($\Pr=\Pm=1$) scaled by $\bra{v_{y0}^2}$ and $\bra{B_{y0}^2}$ respectively.} 
Note the clear exponential growth until $t\approx 1.4 \tAN$. 
Fast oscillations in $\bra{U^2}$ until $t\approx \tAN$  indicate
g--modes originating from the initial velocity perturbation.
{\em Middle panel:} rms values of generated magnetic field components for different runs. 
For legend see lower panel.
Prandtl numbers indicated  
as ($\Pr$,$\Pm$). {\em Lower panel:} rms value of $B_y$.} 
\end{figure}

We first show the temporal evolution of the magnetic field for
a few representative cases in Fig.~\ref{fig:energetics}. 
In all of them, we can clearly distinguish a first stage of exponential growth, 
from a subsequent saturation phase.
The $x$ and $z$ components of the magnetic field are generated
at the expense of its $y$ component.
Although there exists a persistent energy source in the form of a constant heat flux into the domain,
the final saturated stage always undergoes a slow decay. 
This decay is most clearly visible in $B_y$. 
Thus the instability is not able to maintain a dynamo on its own.

We suppose  that the magnetic layer formed by $B_x$,
though having a vertical scale suited to maintain
the instability, is eventually not strong enough to take over the role of the initial magnetic layer. 
Let us first discuss the initial linear stage of the instability.

\begin{figure}[b]
\includegraphics[width=\columnwidth]{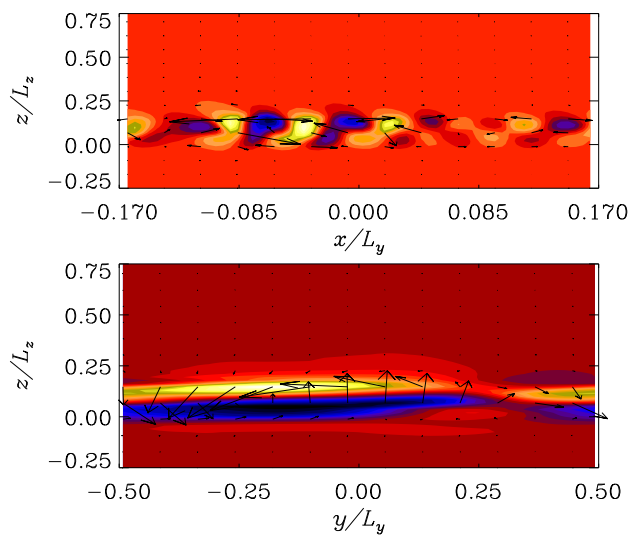}
\caption{\label{fig:lineigmode} 
Top: Velocity components $U_y$ (in color), $U_x$ and $U_z$ (vectors) in the plane $y=0$.
Bottom: $U_x$ (in color), $U_y$ and $U_z$ (vectors) in the plane $x=0$. Both during the 
linear evolution phase of the run B128a.} 
\end{figure}

\subsection{Linear stage.} 
At first we
 verify that the
instability is indeed driven by magnetic buoyancy.
As the coefficients in Eqs. \eqref{eq:continuity}--\eqref{eq:entropy}
are constant, the initial state \eqref{polytr} depends only on $z$,
and the boundary conditions in the $x$ and $y$ directions are periodic,
all eigensolutions 
$\bm{\psi} \in \left\lbrace\rho,\vec{u}, \vec{b},s\right\rbrace$ of
the linearized problem must have the form
\begin{equation}
\label{eq:psi}
\bm{\psi}(\vec{x},t) = \ithat{\bm{\psi}}(z)\, \mathrm{e}^{\displaystyle\iunit \left[2\pi (m x/L_{\rm x}+ n y/L_{\rm y})-\omega t\right]},
\end{equation}
where $m$ and $n$ are integers and
$\omega=\omega_{\rm R}+\iunit\omega_{\rm I}$.
Corresponding dispersion relations $\omega(m,n)$
have been established by applying
perturbations of the form \eqref{eq:psi}
with a variational principle in the non-rotating case \citep{Fan01}
and with the set of linearized anelastic MHD equations     
in magnetostrophic approximation at finite angular velocity \citep{Sch85}.
The former case allows both oscillatory and non-oscillatory unstable modes,
although in \cite{Fan01} only non-oscillatory modes are reported.
In the latter case, however, all unstable modes turn out to be oscillatory with the ratio 
$\omega_{\rm R}/\omega_{\rm I}$ decreasing with latitude.   
Note that the analytic results of \cite{Sch85}
are limited in their predictive power by the fact that the variables are not subjected to our specific boundary conditions
and that the analysis is performed locally. 

For the runs in \Tab{tab:res1} we find 
that in the early exponential growth phase
$m=8$ and $n=1$ throughout as seen in Fig.~\ref{fig:lineigmode}
which shows a typical velocity pattern at a time during the linear stage.

This is consistent with the findings of \cite{Fan01}
where the fastest growing mode had always the smallest possible
(non-vanishing) wavenumber in the direction of the field
whereas the wavenumber perpendicular to the field was high. 
According to the terminology of Fan we may qualify our eigenmodes as {\em undular} as they change 
periodically in the direction of the magnetic background field.
In our case, there seems to be some mixing with lower $m$ modes
since the
growing
perturbations
do not appear to be perfectly sinusoidal. 
While the growth rates presented in \Tab{tab:res1}
could be easily identified from the averaged quantities shown in 
Fig.~\ref{fig:energetics}, it was difficult
to access the oscillation frequencies. This is because they are small compared to the 
growth rates and saturation sets in too early to allow for the observation
of a complete oscillation period.
Nevertheless, some indications for temporal variations in the eigenmode geometries have been found.

\begin{figure}
\hspace*{-.5cm}{\includegraphics[width=0.5\textwidth]{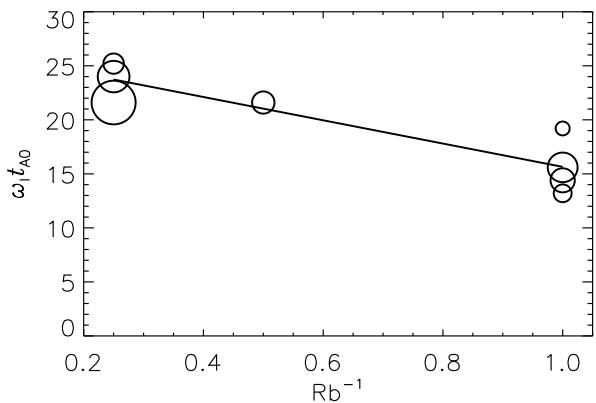} }
\vspace*{-.3cm}
\caption{\label{fig:prm}
Dependence of growth rate $\omega_{\rm I}$ on the inverse Roberts number 
derived from the runs in \Tab{tab:res1}.
Solid line: best linear fit.  
Size of circles codes for the value of $\Rey$ (length-based, see Tab.~\ref{tab:res1}).  
}
\end{figure}

Generally, we observe an increase of the growth rate with increasing magnetic Prandtl number, but a decrease with increasing Prandtl number. 
We find that the growth rate increases with the
Roberts number as shown in Fig.~\ref{fig:prm}.
This means that increasing efficiency of heat conduction in 
comparison to magnetic 
diffusion destabilizes the
sub-adiabatic 
stratification in the system
in agreement with the destabilizing effect of 
thermal diffusion studied by 
\cite{Ach79}.

\subsection{Dependence on initial magnetic field and rotation}

\begin{figure}
\includegraphics[width=0.5\textwidth]{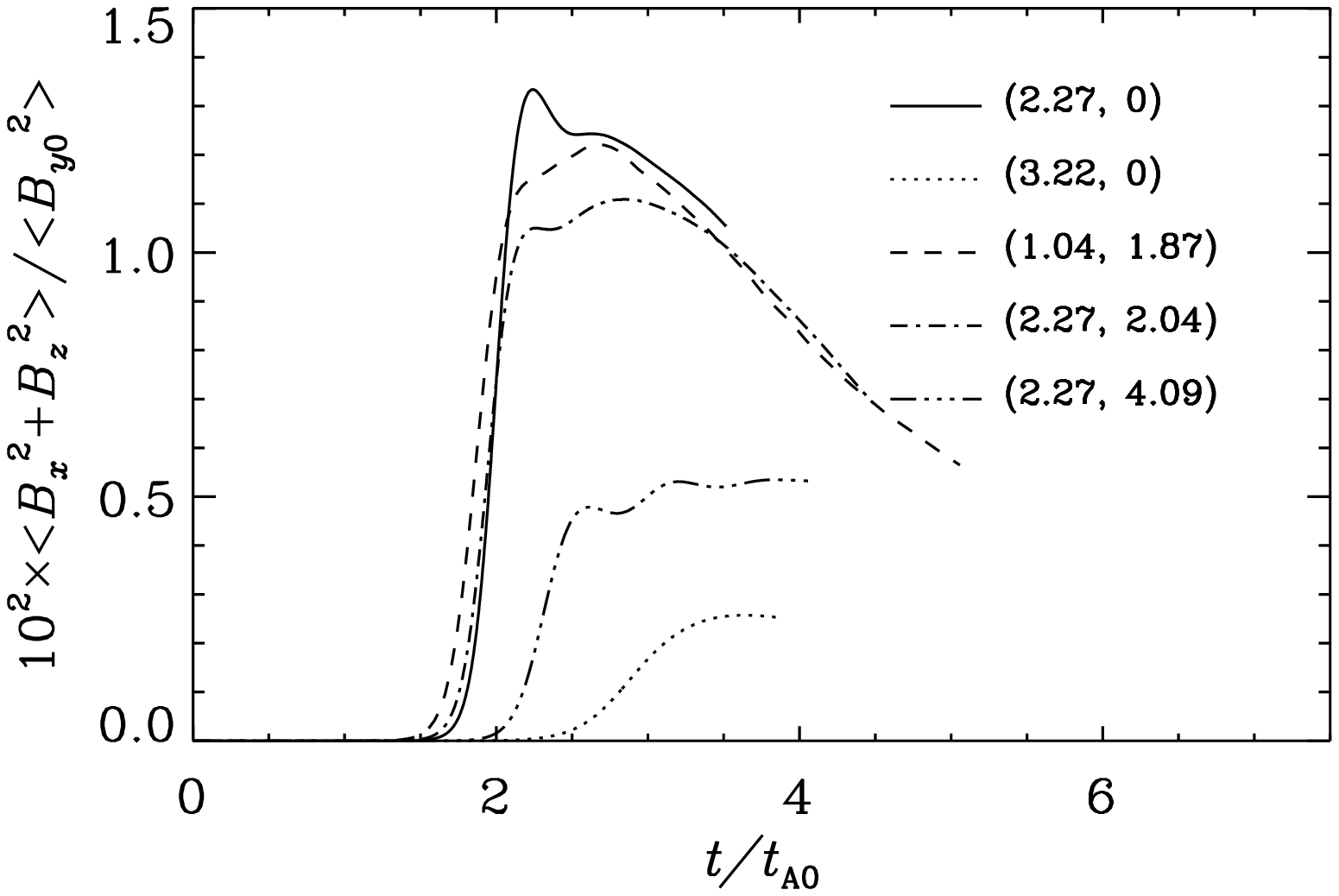}
\includegraphics[width=0.5\textwidth]{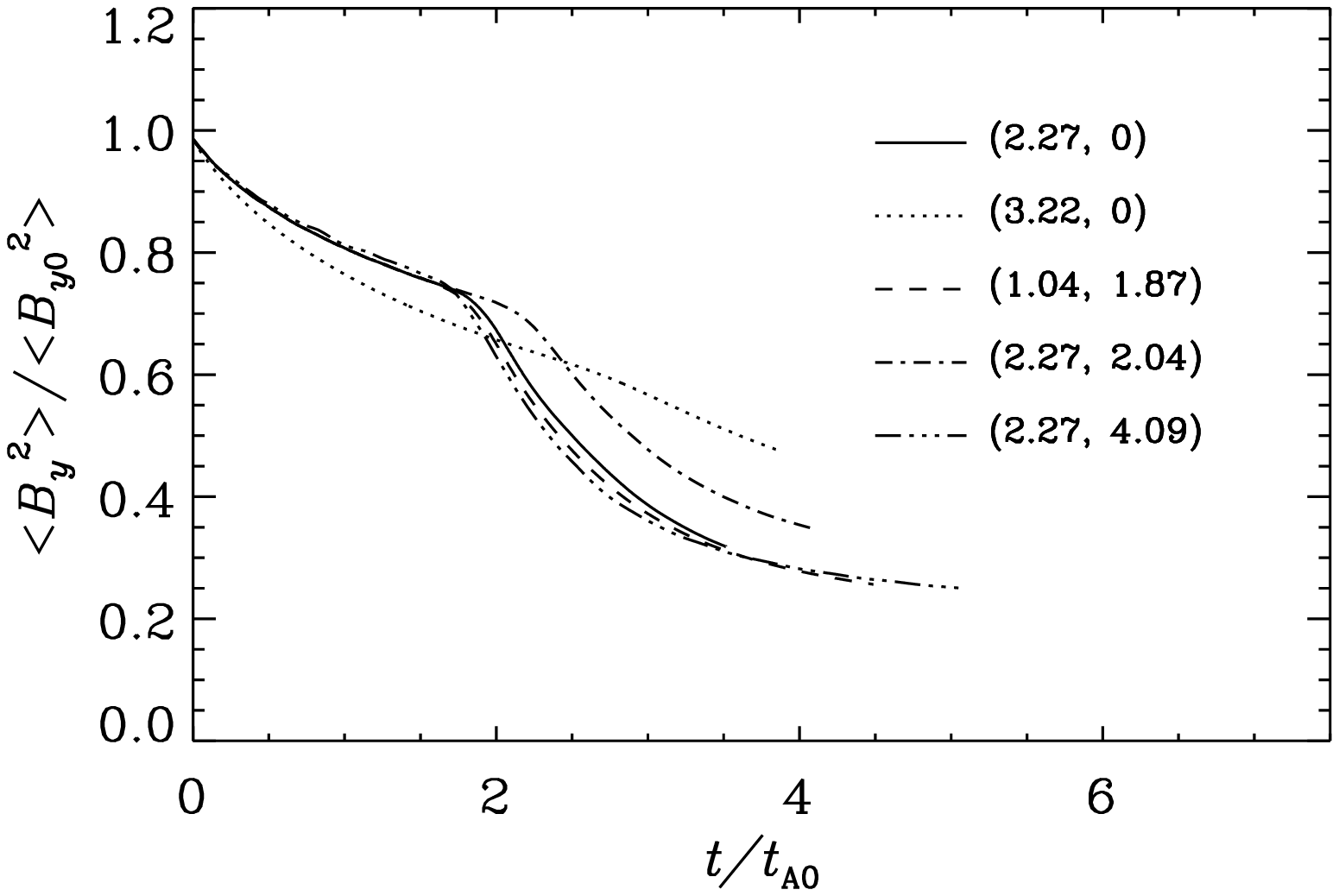}\\[-5.45cm]
\hspace*{.736\columnwidth}$\scriptstyle\boldsymbol{(\betaZ,\:\:\,10^{-5}\zeta)}$\\[5.1cm]
\caption{\label{fig:omgdep} Dependence of the instability
on initial magnetic field strength, expressed by
$\betaZ$ and rotation, expressed by
$\zeta=\betaZ\Tm^{1/2}$.
{\em Upper panel:} rms value of the  
generated magnetic field components
$\bra{B_x^2+B_z^2}$.
{\em Lower panel:}
$\bra{B_y^2}$.
Legend shows $(\betaZ,10^{-5}\zeta)$.
$\Pr=\Pm=1$,   
colatitude $\theta=30^\circ$, resolution $64^3$ throughout.
Note that the normalization time $\tAN$ is not the same for all curves, but proportional to $\betaZ^{1/2}$.
}
\end{figure}

Another piece of evidence for the magnetic character of the instability
is its dependence on the
initial magnetic field strength.
From \figref{fig:omgdep} we see a clear increase of
the
 growth rate
and saturation level with decreasing $\betaZ$,
that is, increasing  $\Lu_0$,
while keeping the rotation rate fixed at $\Ta_{\rm M}=3.24\times10^{10}$.
\cite{Sch00} predicted 
a growth rate $\propto v_{A0}^2/\Omega$ for finite rotation,
in the magnetostrophic approximation,
inversely proportional to $\zeta=\betaZ\sqrt{Ta_{\rm M}}$.

Next we keep $\betaZ$ constant at 2.27 and 
decrease
$\zeta$ gradually from $4.09\times10^5$ to 0.
Inspecting Fig.~\ref{fig:omgdep}, we find that growth rate and saturation
level of $\bra{B_x^2+B_z^2}$ increase monotonically and reach their maxima at
$\zeta=0$ ($\Omega=0$)
while the saturation time is decreasing.
The impeding effect of rotation onto the instability at large $\Omega$ is plausible in view of
the Taylor-Proudman theorem because the unstable eigenmodes
do show pronounced $z$ gradients in $\UU$, see Fig.~\ref{fig:lineigmode}.

\setlength{\textfloatsep}{20pt plus 3pt minus 10pt}

\subsection{Saturated stage.} 
At later time the instability reaches saturation,
characterized by turbulent magnetic, velocity, density and temperature fields,
that decay slowly thereafter.
However, in most of the analysis below, this decay will be ignored and the turbulence approximately 
statistically
stationary. 
The turbulence is necessarily both inhomogeneous and anisotropic and 
we shall further show that it is also helical.
Under such conditions we expect the emergence of a mean electromotive force.
Indeed  magnetic fields perpendicular to the initial magnetic 
layer are produced
having non-vanishing horizontal averages. 

\setlength{\textfloatsep}{0pt plus 3pt minus 20pt}
\begin{figure*}
\vspace{4mm}
\centering{$\Pr = 0.125$, $\Pm = 0.125$, $\Rey= 8.2$}\\[2mm]
\hspace*{-1mm}\includegraphics[width=\columnwidth]{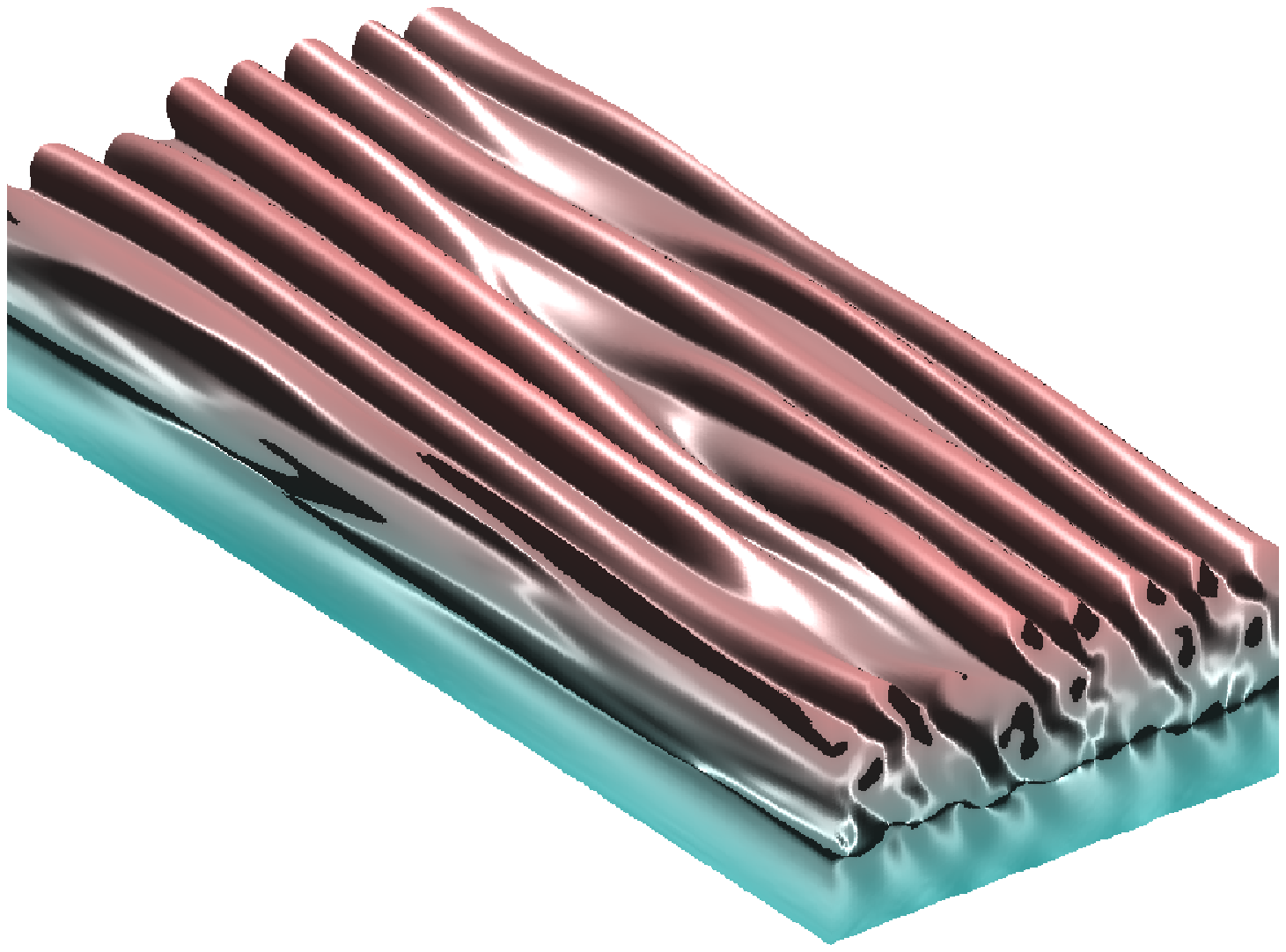} \hspace*{2mm}
\raisebox{10mm}{\includegraphics[width=\columnwidth]{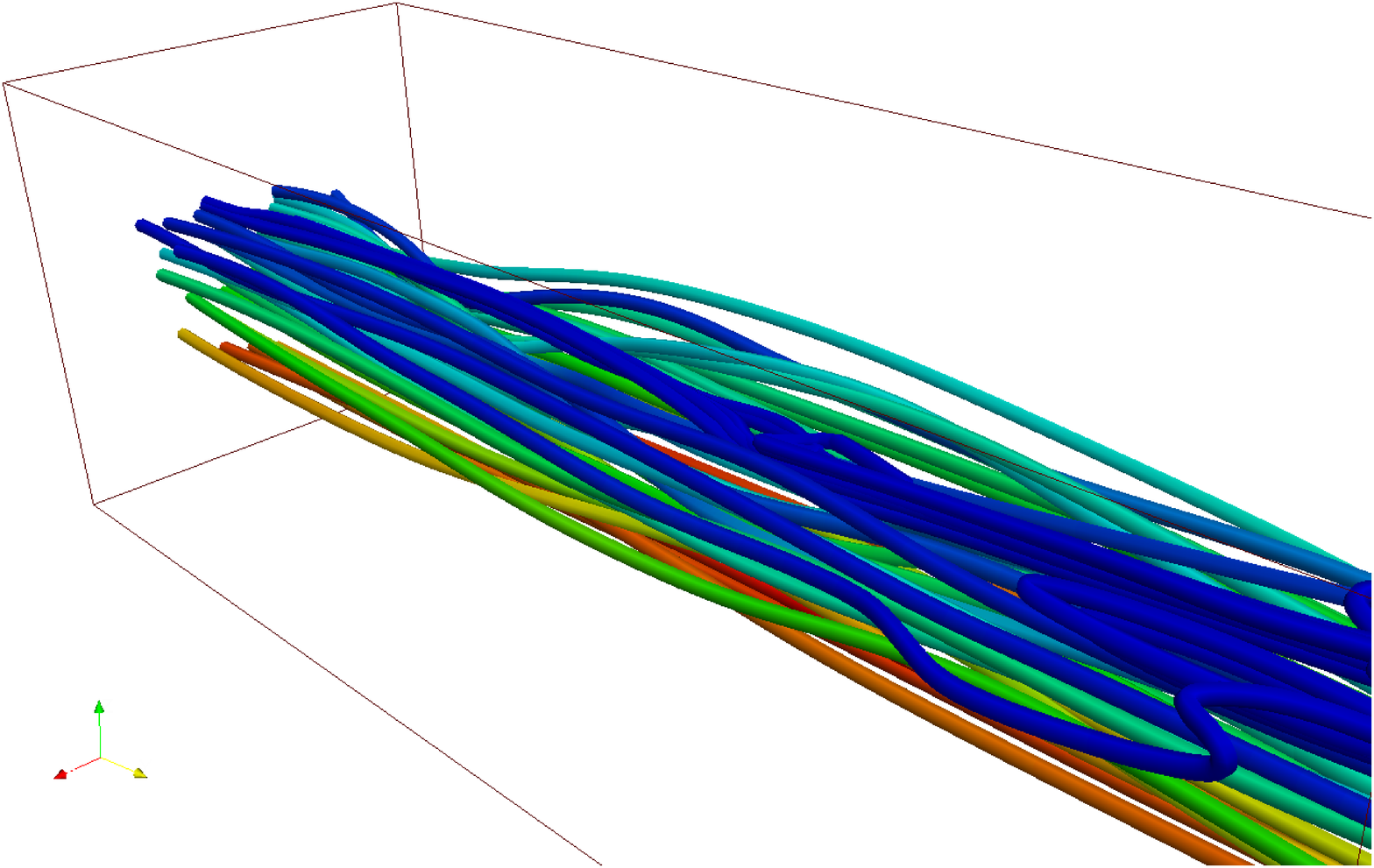}}\\[5mm]
\centering{$\Pr = 0.125$, $\Pm = 0.5$, $\Rey= 3.6$}\\[2mm]
\hspace*{-1mm}\includegraphics[width=\columnwidth]{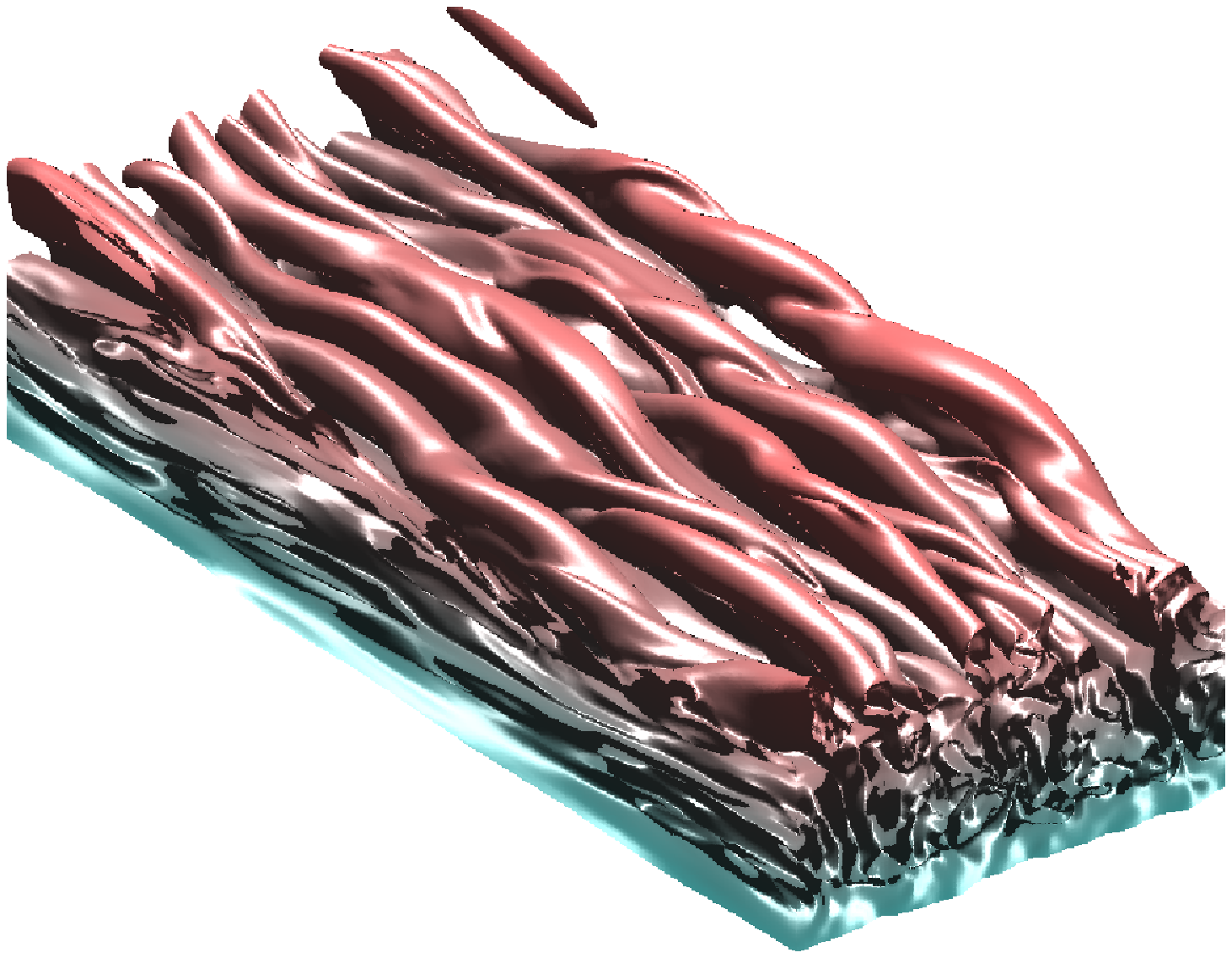}  \hspace*{2mm}
\raisebox{12mm}{\includegraphics[width=\columnwidth]{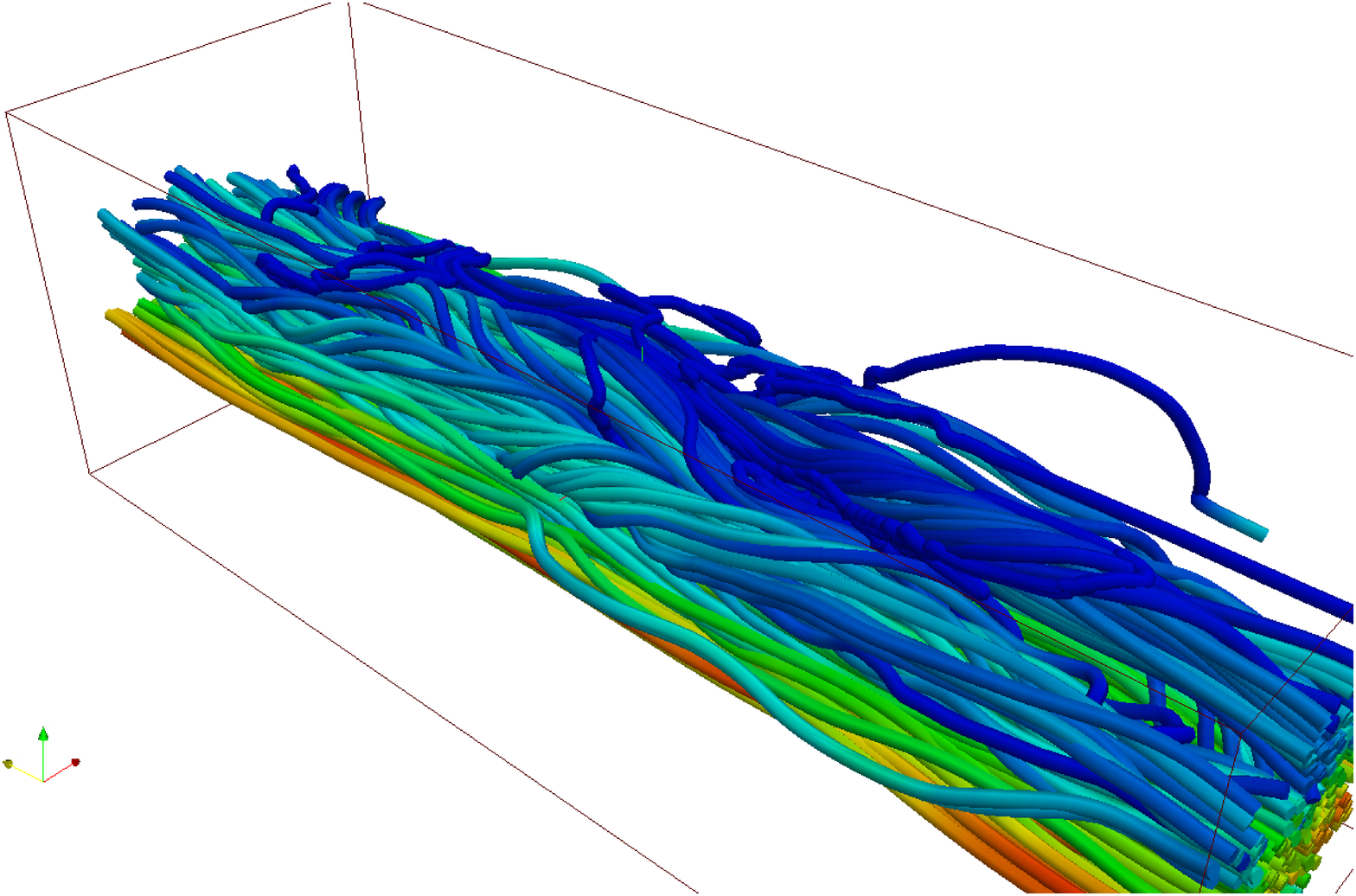}}\\[5mm]  
\caption{\label{fig:volren} Left:
volume rendering of the $B_y=0.1B_0$ isosurface.
Right: field lines, colored according to the value of $B_y$
for runs B128f (top) and B128e (bottom)
at $t=2 \tAN$ (saturated stage).}
\end{figure*}

In order to give a better idea of the 3D geometry of the magnetic field
we provide in Fig.~\ref{fig:volren} a volume rendering of $B_y$ at a
time after $t^\sat$ for the runs B128e and B128f
(see Table~\ref{tab:res1}) which differ only in their magnetic Prandtl numbers.
Notice how the magnetic layer breaks into 
flux tubes  -- similar to what is seen in Fig.~3 of \cite{Fan01}
and also in \cite{MHP95}.
The difference between the two cases is most striking
in the nature of corrugation in the surface shown.
We attribute the difference to larger twist in the rising tubular
structures in the run B128e compared to B128f which becomes clearly
visible in the field line pictures
also depicted
in Fig.~\ref{fig:volren} (right).
\setlength{\floatsep}{0pt}
\begin{figure}[t]
\hspace*{7.6cm}$\delta\rho / \rho_i$\\[-3mm]
\includegraphics[width=.5\textwidth]{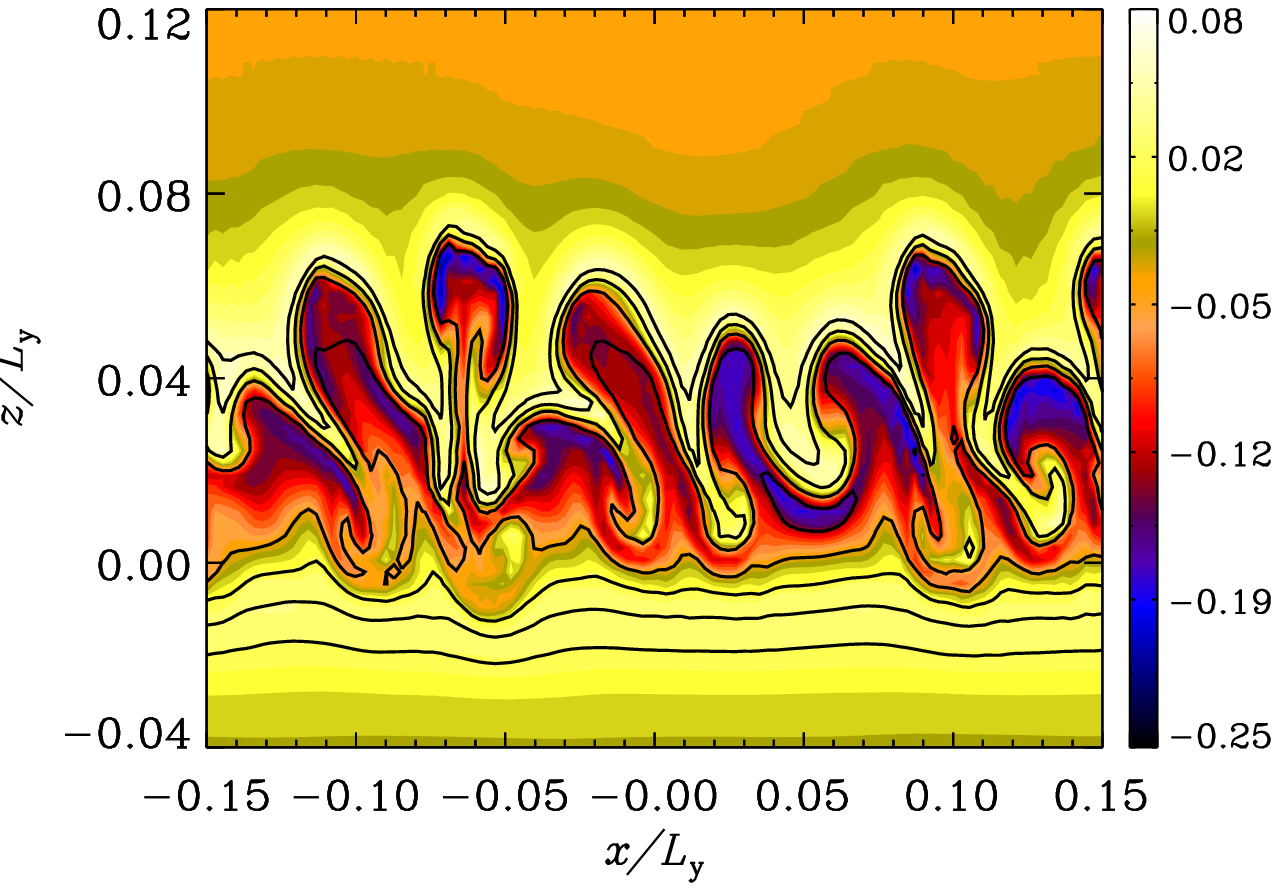}\\[-3mm]
\hspace*{7.6cm}$\delta T / T_i$\\[-3mm]
\includegraphics[width=.5\textwidth]{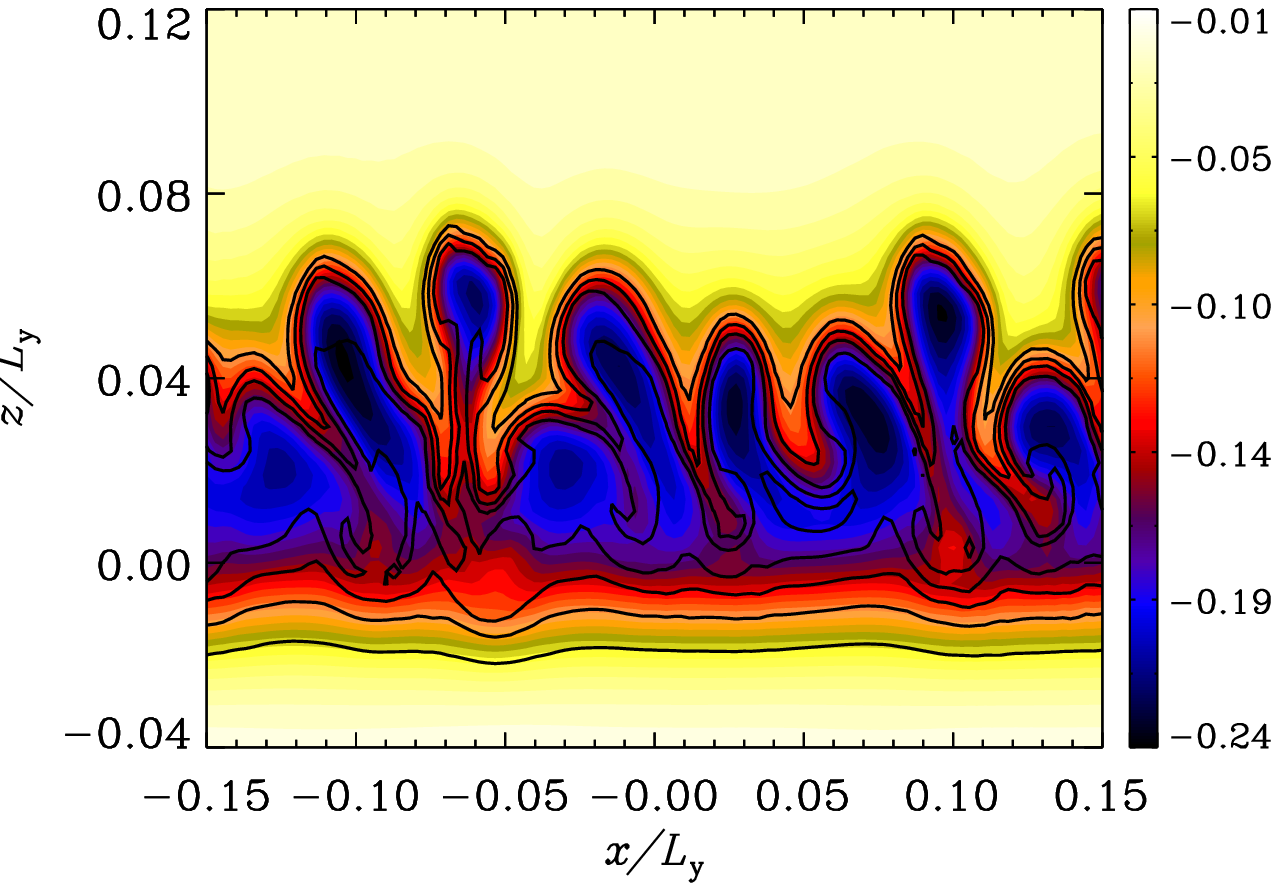}
\caption{\label{fig:den}
Relative density perturbation, $\delta\rho/\rho_i$ (top) and 
relative temperature perturbation $\delta T/T_i$ (bottom), with $\delta\rho=\rho-\rho_i$, $\delta T = T-T_i$ and
$\rho_i(z)$, $T_i(z)$ taken from  Eq.~(\ref{polytr}),
in the plane
$y=0$
at  $t= t^{\rm sat}\sim2\tAN$ for the run B128h.
Both plots  overlaid with contours of $B_y$ (solid lines).
}
\end{figure}
\begin{figure}

\vspace{-.63cm}
\hspace*{-5mm}\includegraphics[width=1.05\columnwidth]{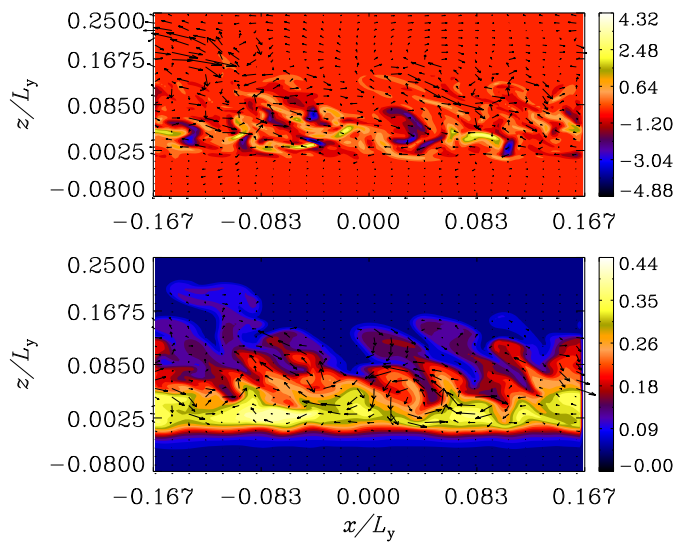}\\[-5mm]
\caption{\label{fig:hel} Scaled current helicity ${\bm J}\cdot{\bm B}/J_{\rm rms}B_{\rm rms}$  
for run B128d (top) at $t=t^{\rm sat}\sim 2\tAN$. Arrows show $v_x$ and $v_z$.
 $B_y/B_{y0}$ (bottom) for the same run.  
 Arrows show $B_x$ and $B_z$.
 Both panels show the plane $y=0$.
 }
\end{figure}

\FFig{fig:den} demonstrates 
the breakup of the magnetic layer into tubular structures of concentrated
magnetic field which are also regions of low density, hence rising.
Notice also the high density regions just above and below
these tubular structures.
They show a significantly lower temperature than their
surroundings (Fig.~\ref{fig:den}, bottom). 

Considering the
solar convection zone
it is suggestive to ask to what extent the flux tubes are twisted, 
as their ability to rise over a large distance
depends crucially on this property. 
 For a quantitative measurement
we utilize the dimensionless parameter
$\varepsilon_J=\bra{\JJ\cdot\BB}/J_\rms B_\rms$,
the {\em relative current helicity}, essentially measuring the overall degree of alignment between $\BB$ and $\JJ$.
Here, angular brackets denote volume averages.
A corresponding localized quantity is
$\epsilon_J(z) = \overline{\JJ\cdot\BB}/J_\rms B_\rms$. 
\FFig{fig:hel} shows 
$\varepsilon_J$ (filled contours)
as well as $B_y$ in the plane $y=0$
for run B128e.
Notice that the contours are bend leftward because of the 
Coriolis force with
$\vec{\Omega}\cdot\vec{\hat{z}} > 0$. The contour plots of $B_y$ 
in this figure also show the formation of rising tubular structures 
from the magnetic layer.

\begin{figure}

\centering{\includegraphics[width=.45\textwidth]{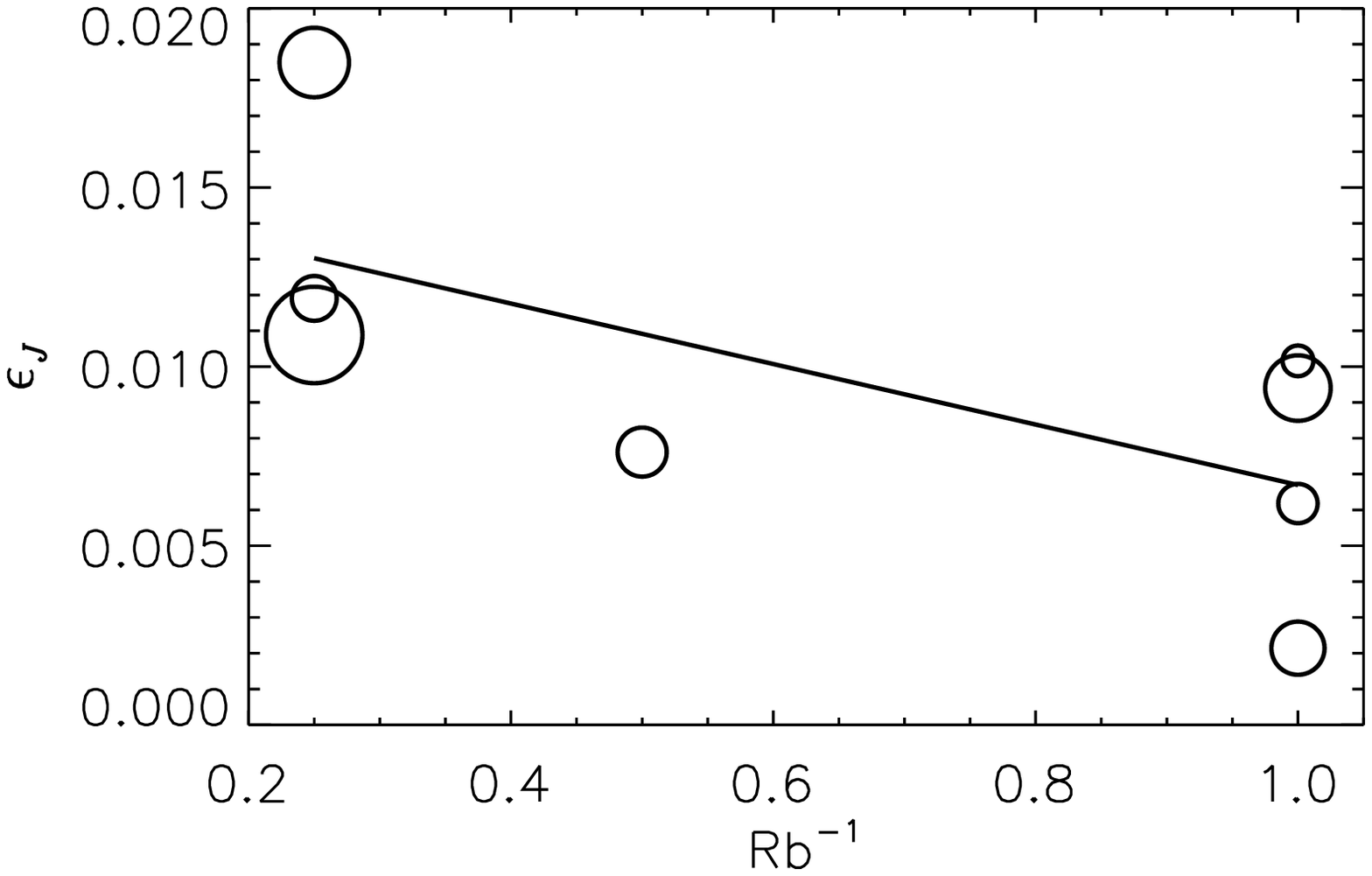}}\\
\centering{\includegraphics[width=.40\textwidth]{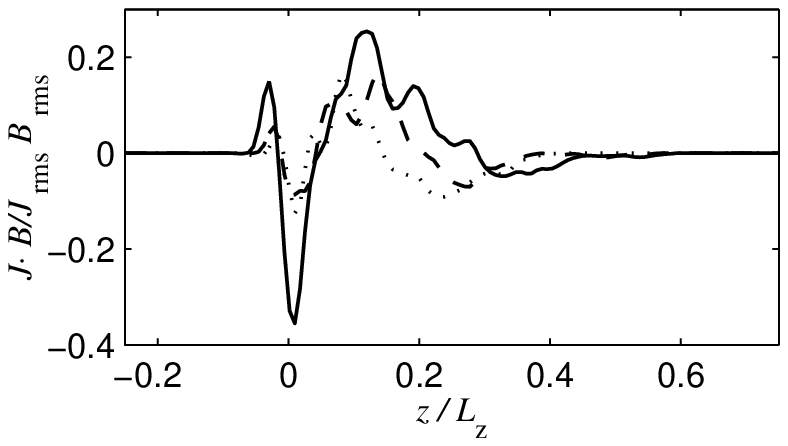}}\\
\caption{\label{fig:prm2}
Top: Dependence of the total relative current helicity $\varepsilon_J$  
on inverse Roberts number.
Size of circles codes for value of $\Rm$.
Bottom: Dependence of $\overline{\vec{J}\cdot \vec{B}}/J_{\rm rms}B_{\rm rms}$    
on $z$ after saturation for runs B128e (solid, $\Rb^{-1}=0.25$), 
B128h (dashed, $\Rb^{-1}=0.5$), and B128c (dotted, $\Rb^{-1}=1$).
}
\end{figure}

In Fig.~\ref{fig:prm2} we show the dependence of $\varepsilon_J$ on $\Rb$ and 
the profiles $\epsilon_J(z)$ for some selected runs. 
Although the total helicity reaches only values of a few percent, its localized counterpart is as strong as $30 \%$ near to 
the initial location of the magnetic sheet. 
The clear dependence of $\varepsilon_J$ on $\Rb$ is
in contrast to the only weak dependences on
$\rm Pr_{\rm M}$ and $\rm Pr$ individually. This is an important result
from this section. 
Our conjecture is that, at large $\Rb$, this magnetic buoyancy instability 
may play an important role in the formation of twisted flux tubes 
in the Sun, where $\Rb \gg 1$ is expected.
\setlength{\floatsep}{20pt}
\begin{figure}[b]

\vspace{1mm}
\hspace*{4mm}\includegraphics[width=0.45\textwidth]{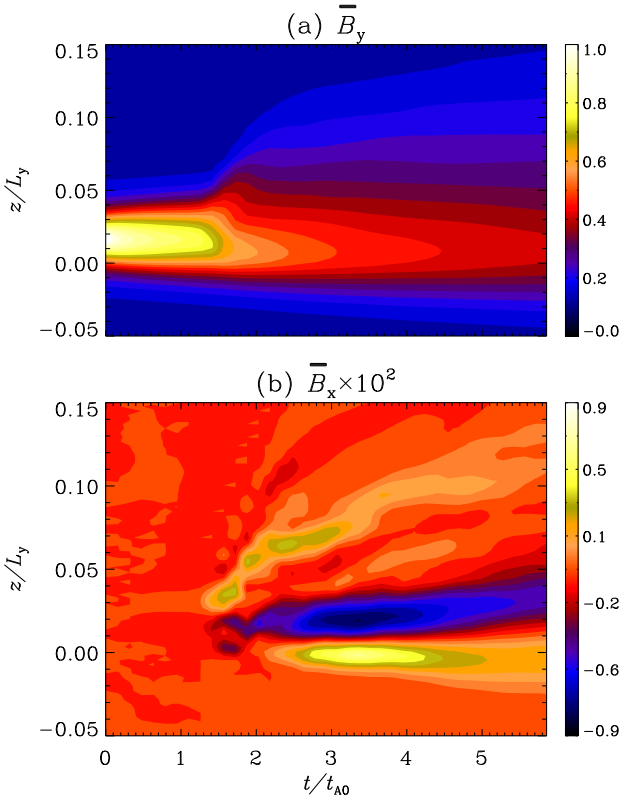}
\caption{\label{fig:meanbxby}
Time-depth 
diagram for $\meanB_{x,y}$ normalized on $B_0$ for 
run B128g in \Tab{tab:res1}
($z$ extent of the box clipped).
Note the difference of two orders of magnitude between $\meanB_y$ and $\meanB_x$.
}
\end{figure}  

To demonstrate the emergence of a mean magnetic field we present in Fig.~\ref{fig:meanbxby}
time-depth plots of $\ob{B}_y$ and $\ob{B}_x$ for the run B128g (note that
$\meanB_z=0$).
There, $t\approx 1.6\,\tAN$ marks the end of the exponential growth phase
after which a strong growth of $\meanB_x$, obviously
at the expense of $\meanB_y$, sets in.
$\meanB_x$ reaches its maximum
around $t \approx 3\tAN$ and
is
then subject to the overall decay.
Note the strong vertical concentration of $\meanB_x$, approximately
antisymmetric about the midplane of the magnetic sheet.

\subsection{Calculation of turbulent transport coefficients}

The turbulence resulting from the buoyancy instability generates
a mean magnetic field component $\meanB_x$ from an initial $\meanB_y$
which
is also 
modified
compared to its initial shape
 (see Fig.~\ref{fig:meanbxby}).
It is then natural to employ the technique 
known as the
{\em quasi-kinematic test-field method} to 
calculate transport coefficients like the $\aTens$ and $\eTens$ 
tensors which describe this process.
So far, test-fields have mostly been used in situations where a 
hydrodynamic background 
was 
already present
in absence 
of the mean magnetic field  \citep[see, e.g.] []{BRS08,BRRK08,BRRS08}.
Here, in contrast, the (magnetohydrodynamic)
turbulence results entirely from the instability of a  
pre-existing mean magnetic field, $B_{y0}(z)$.
In other words, our simulations do not posses a kinematic stage
in which the influence of $\meanBB$
would be
negligible.
One might worry that in such a situation the quasi-kinematic
test-field method fails \citep{Courvoisier10}.
However, Eq.~\eqref{eq:dbdt} continues to be valid and hence all
conclusions drawn from it, because the decisive
applicability
criterion is whether or
not there exists
hydromagnetic
turbulence in the absence of the mean magnetic field.
This is not the case here,
so the method should be applicable.
The only peculiarity occurring is the fact that
all components of $\aTens$ and $\eTens$
vanish for $0\le \meanB_\rms \le \meanB_{\mathrm{threshold}}$, because
fluctuating velocity and magnetic fields develop only after the
instability has set in.
Another aspect not considered in most previous test-field studies is
the strong intrinsic inhomogeneity of the turbulence not only 
as a consequence of the strong $z$ dependence of $\meanBB$, 
but also due to the stratified density background.
Thus the transport coefficients need to be 
determined
as $z$ dependent quantities.
We shall next demonstrate that the test-field method
still works reasonably well in this regime.
Note that to calculate the transport coefficients in addition
to the usual MHD equations
four additional
evolution equations of 
the form \eqref{eq:dbdt}
for  four independent test-fields
have to be solved.
Hence the test-field runs are computationally almost thrice as expensive.
We have thus reduced resolution to $64^3$ grid points for all
these runs.

\subsubsection{Reconstruction of the mean EMF}

To validate the test-field method we first
confirm that the quantity $\meanEMF$,
taken directly from the DNS,
can be reproduced by employing the relation
\eqref{eq:Eansatz}
between  $\meanEMF$ and $\meanBB$
with the
tensors $\aTens$ and $\eTens$ 
determined using the quasi-kinematic test-field method. 
In mathematical terms,
\begin{align}
\hspace*{-16mm}\meanemfs_i^{\rm R}(z; \meanBB)\!=
\!\sum_{k}\!\big[\hat{\cal K}^{\crm}_{ij}(z,k; \meanBB)
\ithat{B}^{\crm}_j(k) \label{assembly}    
+\hat{\cal K}^{\srm}_{ij}(z,k; \meanBB) 
\ithat{B}^{\srm}_j(k) \big]   
\hspace*{-1cm}
\end{align}
with
\begin{align}
&\begin{alignedat}{2}
\hat{\cal K}_{ij}^\crm &=\hat{\alpha}_{ij}(z,k^\crm; \meanBB)\cos(k^\crm \itilde{z}) &&- \hat{\eta}_{il}(z,k^\crm; \meanBB)\epsilon_{lj3} k^\crm \sin(k^\crm \itilde{z}), \\
\hat{\cal K}_{ij}^\srm &=\hat{\alpha}_{ij} (z,k^\srm; \meanBB)\sin(k^\srm \itilde{z}) &&+ \hat{\eta}_{il}(z,k^\srm; \meanBB)\epsilon_{lj3} k^\srm \cos(k^\srm \itilde{z}), \\
\end{alignedat}\nonumber \\[1mm]
&\begin{aligned}
\ithat{B}^{\crm}_j(k) &=  \frac{2}{L_z} \int_z \meanB_j(z)\cos(k^\crm \itilde{z})\, \dd z, \\
\ithat{B}^{\srm}_j(k) &=  \frac{2}{L_z} \int_z \meanB_j(z)\sin(k^\srm \itilde{z})\, \dd z,
\end{aligned}\label{eq:spectral}\hspace*{-0cm}\\[2mm]
&\hspace*{1mm}k^\crm = \frac{(2k-1)\pi}{L_z}, \quad k^\srm = \frac{2k\pi}{L_z}, \quad k=1,2,\ldots, \nonumber\\
& \quad \itilde{z} = z - z_0-\frac{L_z}{2},\nonumber
\end{align}
where the superscript R indicates reconstruction.
Here, the boundary condition for $\BB$ gives rise to the selection of
discrete cosine and sine modes with wavenumbers  $k^\crm$ and $k^\srm$,
respectively.
The additional argument $\meanBB$ is to indicate that the kernels
$\boldsymbol{\hat{\cal K}^{\crm,\srm}}$,
as well as the tensors $\aTens$ and $\eTens$,
are valid just for that mean field $\meanBB$ 
which is present in the main run.
As a consequence, the reconstruction of the mean EMF can be successful
only when employing exactly this $\meanBB$ in \eqref{eq:spectral}.
That is, the mean field
representation of the turbulence by $\aTens$ and $\eTens$
has, at this level, merely descriptive rather than predictive potential.

\begin{figure*}
\includegraphics[width=0.48\textwidth]{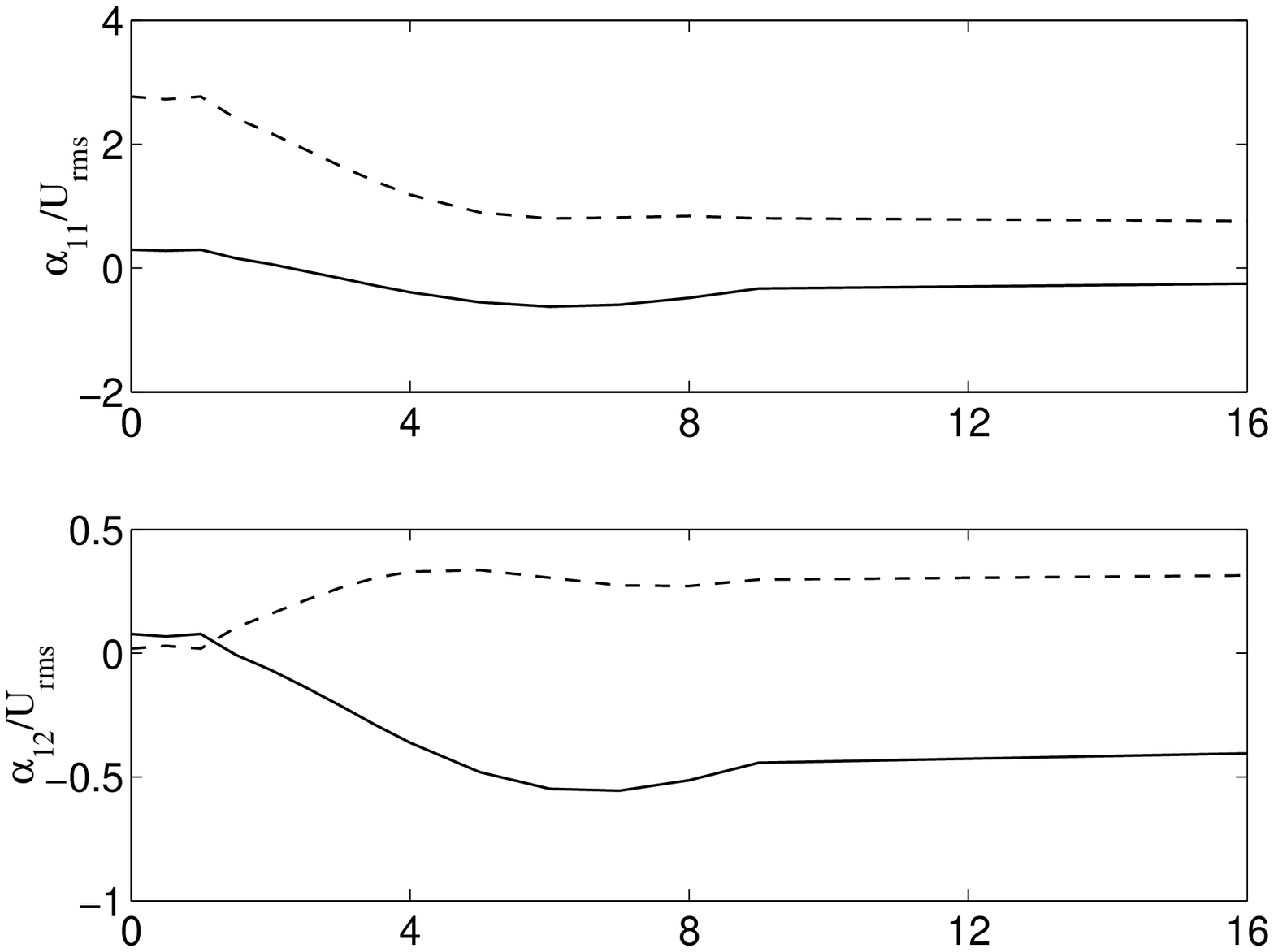}\hspace{3mm}
\raisebox{-2mm}{\includegraphics[width=0.492\textwidth]{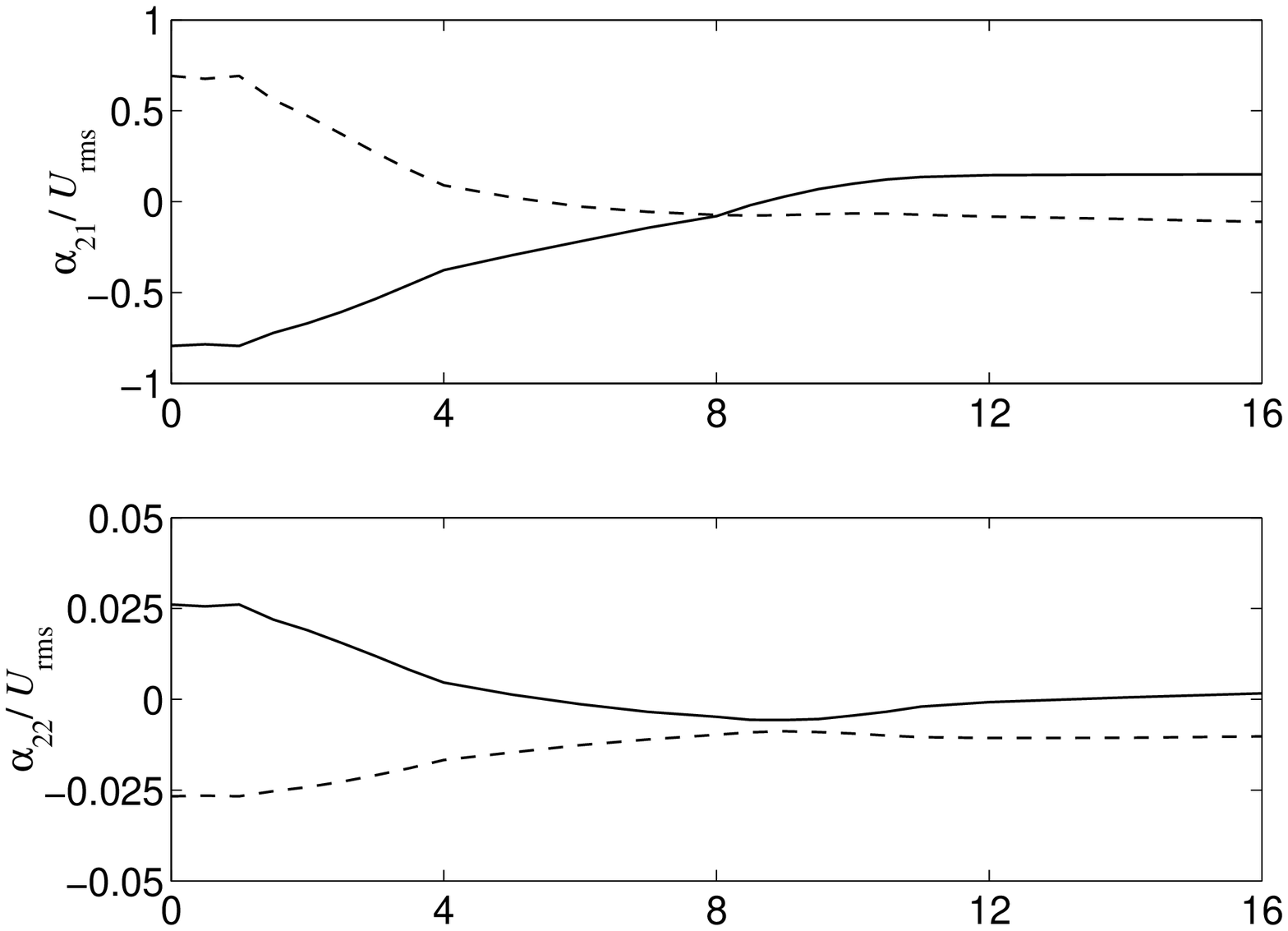}}\\[-2mm]
\includegraphics[width=0.48\textwidth]{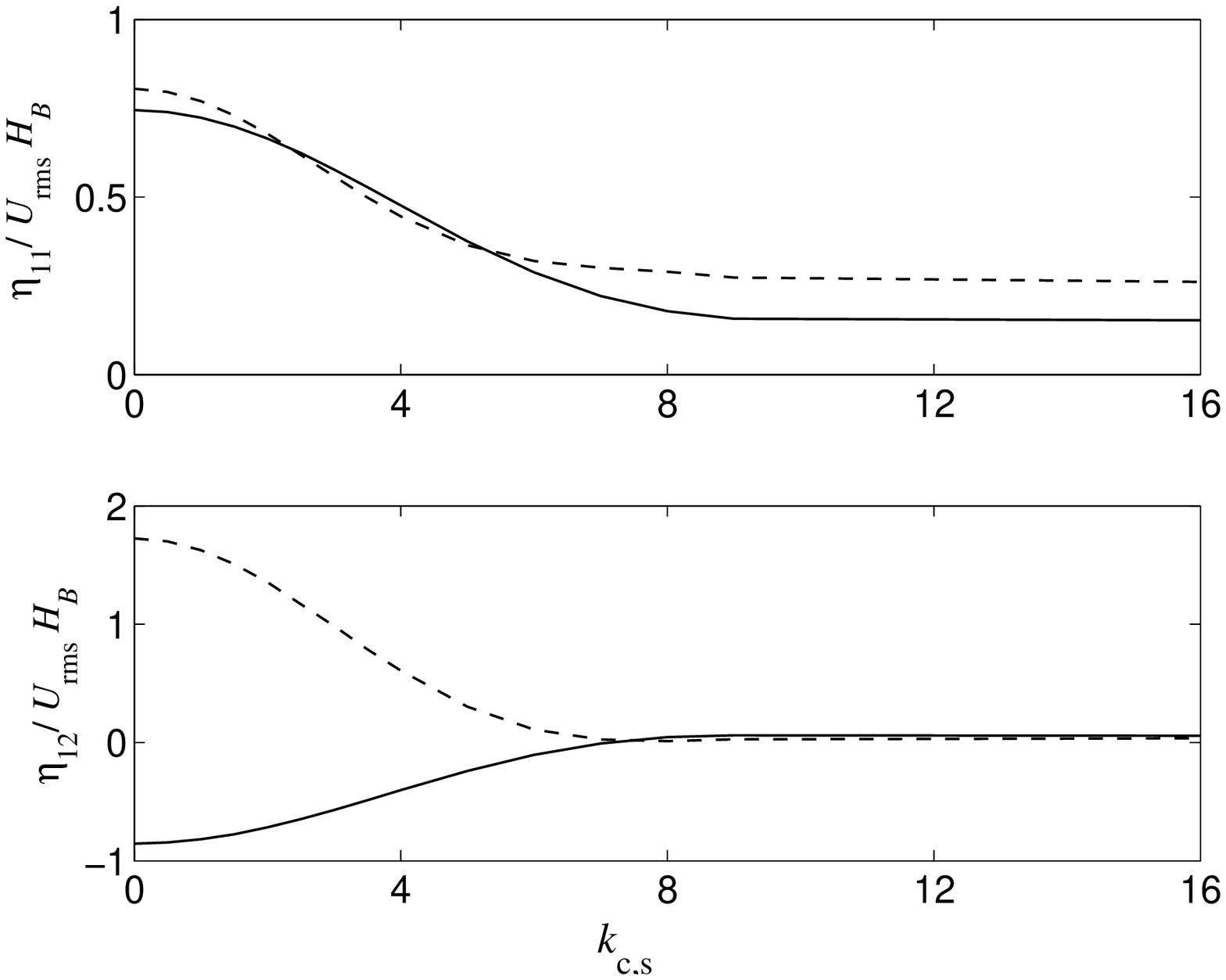}\hspace{3mm}
\raisebox{-2mm}{\includegraphics[width=0.492\textwidth]{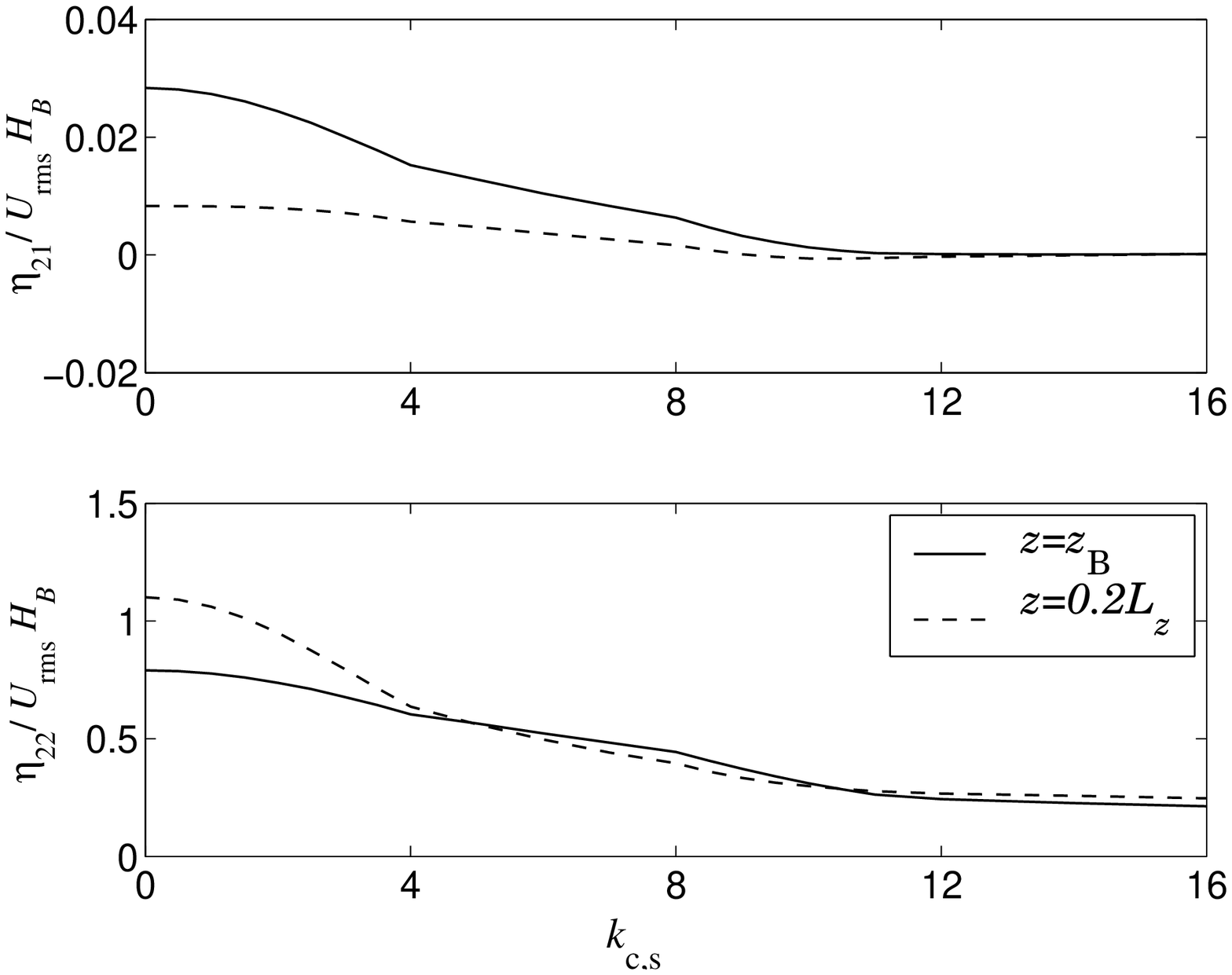}}\\[-2mm]
\caption{\label{fig:kdep} 
Dependence of 
$\aTens$ and $\eTens$
on the test-field wavenumber $k'$
in the midplane of the magnetic layer ($z=z_B$) and near the midplane of the box ($z=0.21 L_z$)
for run TF30+ of Table.~\ref{tab:TF}.
Integer and half-integer values of $k'$ belong to sine and cosine modes in $\meanBB$, respectively.
Note that $k'=0$ refers to constant and linear test fields  and that the coefficients
for that value do not enter the $\meanEEEE$ - $\meanBB$ relation for the given setup.
}
\end{figure*}

\begin{figure}[t]
\hspace*{0.01\columnwidth}$\meanemfs_{x}$\\[-0.8cm]
\hspace*{-1.5mm}\includegraphics[width=0.517\textwidth]{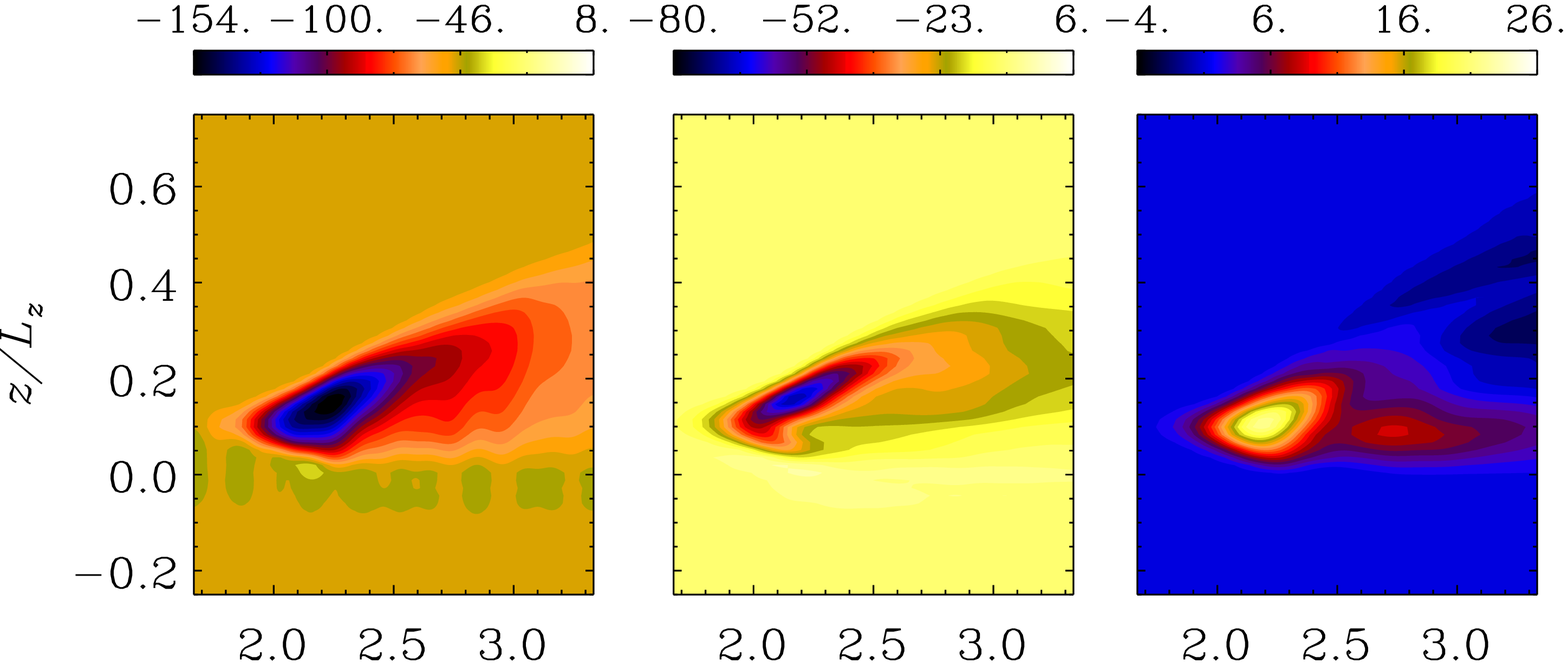}\\
\hspace*{0.01\columnwidth}$\meanemfs_{y}$\\[-0.8cm]
\hspace*{-1.5mm}\includegraphics[width=0.517\textwidth]{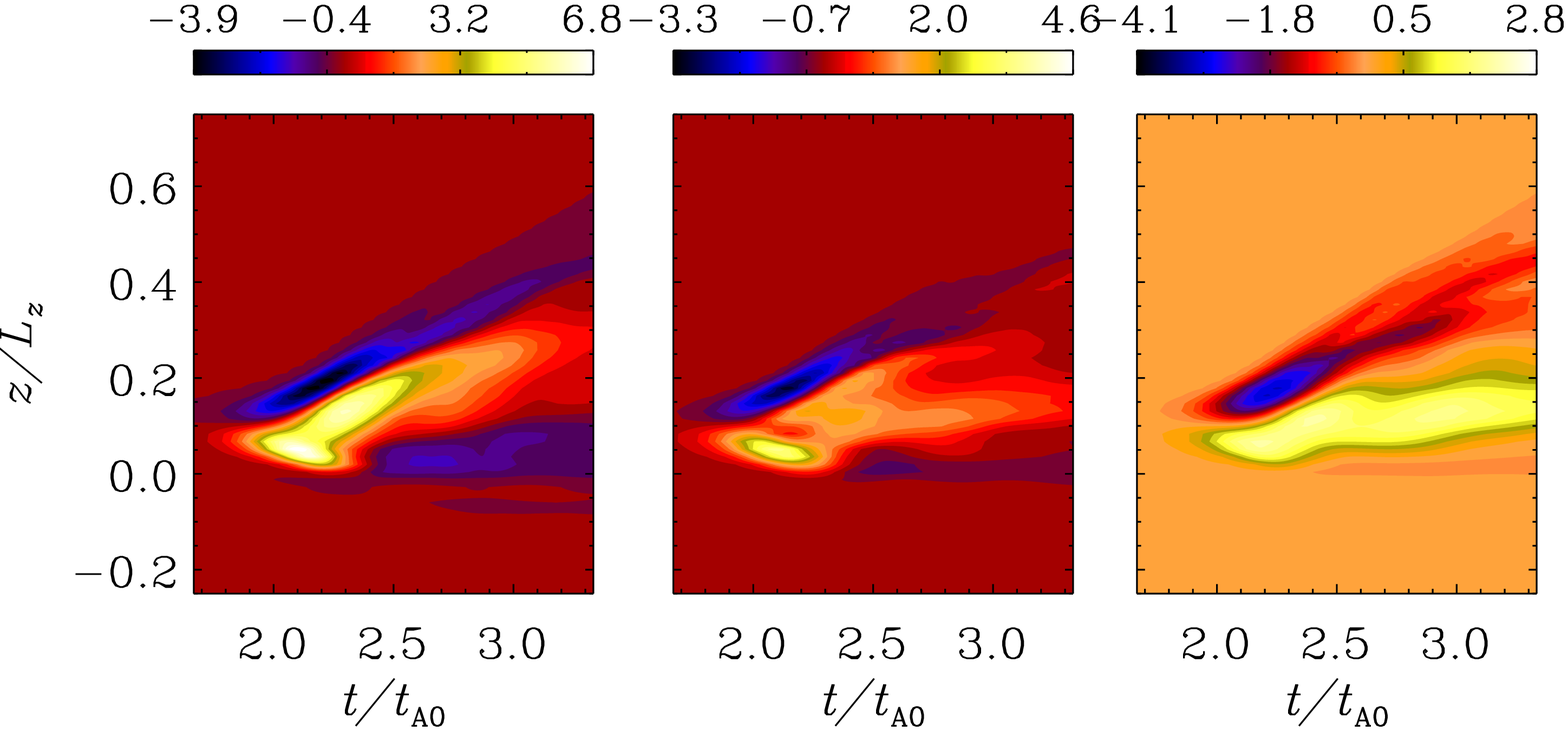}
\caption{\label{fig:emfconst} 
Reconstruction of the mean EMF for the run TF30+ using
$\aTens$ and $\eTens$ from the  test-field method. 
Top: $\meanemfs_x(z,t)$, bottom: $\meanemfs_y(z,t)$, both scaled by $10^{-4}\vAN B_0$.
Left: directly
from $\overline{{\bm u}\times {\bm b}}$.
Middle: Reconstruction
using  all contributions $k'=0.5,1, 1.5, ...16$ in (\ref{assembly}).
Right: Same as before, but using only the $k'=0.5$
contribution.
}
\end{figure}

\begin{figure}
\includegraphics[width=0.95\columnwidth]{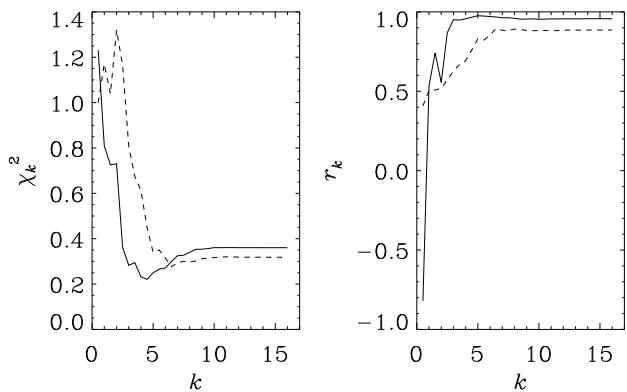}
\caption{\label{fig:stat}
Quality of the EMF reconstruction
as a function of the truncation wavenumber $\kmax$:
  $\chi_{k'}^2$ (left) and correlation $r_{k'}$ (right) 
calculated for $\meanemfs^{\rm R}_x$ (solid) and $\meanemfs^{\rm R}_y$ (dashed) using Eq.~(\ref{eq:goodness}). }
\end{figure}

Let us denote 
$\meanEMF^{\rm R}$ 
as the reconstructed EMF according to 
Eq.~(\ref{assembly}) truncated at
$k' \le \kmax$, with
$k^{c,s} = 2k'\pi/L_z$. 
Here $k'$ can now take both integer and half-integer values
where
the integer (half-integer) 
values of $k'$ correspond to the family of sine (cosine) modes in Eq.~(\ref{eq:spectral}).
An initial estimate of $\kmax$ required for a 
reasonable reconstruction of $\bm \meanemfs$ was obtained from the 
power spectra of both $\meanB_x$ and $\meanB_y$. It turned out that $\meanB_x$ has significant 
spectral power 
up until $k' = 16$, whereas for $\meanB_y$ the power spectra has
levelled off already at $k'=8$.
The components of the tensors $\aTens$ and $\eTens$ also show rather
different spectral behavior, 
both in the
midplane of the magnetic layer and near the midplane of the box as seen in \figref{fig:kdep}. 
From the figures it is evident that
in most cases the spectra can be reasonably truncated at $k'=8$
with the exceptions of $\alpha_{12}$ and $\alpha_{21}$.
Note that the values for $k'=0$ are not relevant here as, due to the boundary conditions,  
$\meanBB$ does not possess a $k'=0$ contribution.
The result of the assembly of $\meanEMF^{\rm R}$
from
 \eqref{assembly} with
\eqref{eq:spectral}, is presented in
Fig.~\ref{fig:emfconst}, middle column.
From simple visual inspection  we find it to be a faithful reproduction
of $\meanEMF$ from the DNS shown in the left
column.
Clearly, a naive application of the test-field procedure with
harmonic test fields with only the lowest $k'=0.5$ results in an
inadequate description as shown in the
right
column.
We
define two measures 
for the quality of the mean EMF reconstruction
namely
$\chi^2_{k'}$
and the correlation coefficient $r_{k'}$ defined as
\begin{equation}
\label{eq:goodness}
\hspace*{-2.5mm}\chi^2_{k'} = \frac{\bra{(\meanemfs_{x, y}-\meanemfs^{{\rm R}}_{x, y})^{2}}_{z, t}}{\bra{\meanemfs_{x, y}^2}_{z, t}}, \hspace*{1.5mm}
r_{k'} = \frac{\bra{\meanemfs_{x,y}\times\meanemfs^{\rm R}_{x,y}}_{z, t}}
{\sqrt{\bra{\meanemfs_{x, y}^2}_{z, t}\bra{\meanemfs^{\rm R 2}_{x, y}}_{z, t}}},
\end{equation}
where the subscript ``$z,t$" denotes that the averaging has been carried out
over the vertical coordinate $z$ as well as over the temporal range
$1.2\tAN \leq t \leq 3.4\tAN$.
The relative error
of
the reconstruction, $\chi^2_{k'}$,
and the correlation
coefficient,
$r_{k'}$, are plotted in Fig.~\ref{fig:stat} as a function
of the
truncation wavenumber $\kmax$.
The $\chi^2_{k'}$ reach a minimum
value and level off around $k' = 8$ for both $\meanemfs_x$ and $\meanemfs_y$.
This implies that including higher harmonic test fields beyond $k'=7$
does not improve the reconstructed EMF. We speculate that the reason
behind this discrepancy is that we have neglected memory
effects \citep{hub+bra09} in the turbulent transport coefficients.
This can be particularly important
in the present situation as we are obviously not in
a statistically stationary regime. 
Similarly $r_{k'}$ for $\meanemf^{\rm R}_x$ $(\meanemfs^{\rm R}_y)$ converges to a value
of 0.98 (0.93) at $k' = 4$ $(8)$.
It is important to note that even though the tensor components 
$\alpha_{12}$ and $\alpha_{21}$ do not converge with increasing $k'$, the reconstructed
EMFs do. Also calculating transport coefficients for $k' \geq 8$ does not improve 
the reconstruction any further. 
This is probably because we do not sufficiently resolve wavenumber scales larger than 
$10$ in the domain with a grid resolution of only $64^3$.

\setlength{\textfloatsep}{20pt plus 3pt minus 20pt}
\subsubsection{Dependence of the transport tensors on inclination}

From the point of view of the solar dynamo it is important
to look at $\aTens$ and $\eTens$ as functions of the
rotational inclination $\theta$ or
latitude $\lambda$,
with a focus 
on symmetry properties with respect to $\lambda=0$, 
which is the solar equator.
Moving from the northern hemisphere at $\lambda$ to the southern at
$-\lambda$, that is
changing $\theta$ to $\pi-\theta$, but keeping all other problem parameters constant,
is equivalent to inverting the sign of $\Omega_z$.
As the same can be accomplished by reflecting the corresponding rigid rotation
about the plane $x=0$, we might construct the solution $(\rho, \UU, \BB, s)$ of 
\eqref{eq:continuity}--\eqref{eq:entropy} for $-\lambda$
simply by reflecting it properly
about the same plane.
Under this reflection polar vectors like velocity transform as,
\EQ
\label{eq:refl}
\left\lbrace U_x, U_y, U_z\right\rbrace(x,y,z) \rightarrow \left\lbrace -U_x, U_y, U_z\right\rbrace(-x,y,z),
\EN
and axial vectors like the magnetic field as
\EQ
\left\lbrace B_x, B_y, B_z\right\rbrace(x,y,z) \rightarrow \left\lbrace B_x, -B_y, B_z\right\rbrace(-x,y,z)
\EN
(Note that the gravitational acceleration is invariant under this reflection.)
Hence, for the initial magnetic field, $B_{y0}(z)$, the transition to $-\lambda$ 
requires only a sign inversion. 
But, since the
induction equation is linear in $\BB$, 
and Lorentz force
as well as Ohmic dissipation 
are quadratic, inverting the sign
of $B_{y0}(z)$
would just transform the solution $\{\rho,\UU,\BB, s\}$ to  $\{\rho,\UU,-\BB, s\}$, 
that is, would leave the turbulence essentially unchanged and can be omitted.
Moreover, as the transport coefficients,
expressing correlation properties of
the turbulent velocity $\uu$,
are functions of $z$ only 
the reflection operation can hardly change their magnitudes.
With respect to possible sign inversions we consider, that
$\meanEEEE$ and $\meanJJ$, being polar vectors, invert the sign of their $x$ components 
under reflection, but keep their $y$ components unchanged. The axial vector $\meanBB$ behaves 
just the opposite way. Thus, we have $\alpha_{ii} \rightarrow -\alpha_{ii}$ for $i=1,2$ (no summation) and
$\alpha_{ij} \rightarrow \alpha_{ij}$ for $i\ne j$, whereas $\eta_{ii} \rightarrow \eta_{ii}$ for $i=1,2$ and
$\eta_{ij} \rightarrow -\eta_{ij}$ for $i\ne j$ when moving from $\lambda$ to $-\lambda$.
Consequently,
it appears that the results for the
southern hemisphere can be derived from those for the northern by simple operations.
Strictly speaking however, this is only true when the initial condition for 
$\UU$ is also reflected upon the transition from $\lambda$ to $-\lambda$.
From a naive point of view we might suppose that omitting this reflection can hardly be of any 
importance, 
because we use random initial condition.
But this we have found not to be true.
We note further that once the initial condition is reflected too
the symmetry is restored. 

\begin{figure}
\includegraphics[width=0.5\textwidth]{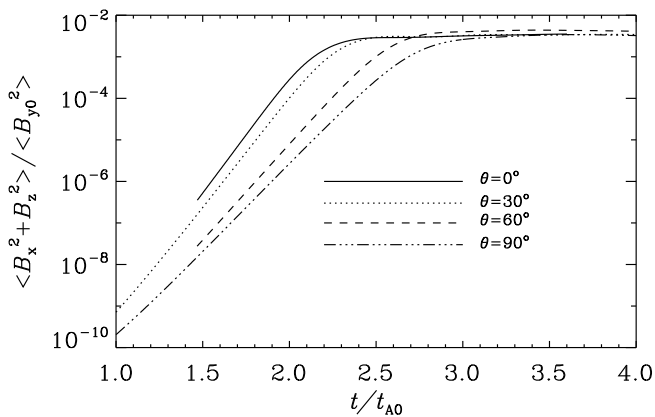}
\caption{\label{fig:thetadep} Dependence of the instability on
rotational inclination $\theta$ in terms of
rms value of generated field components  $\bra{B_x^2+B_z^2}$ for the runs 
TF$0+$, TF$30+$, TF$60+$ and TF$90+$ in Table.~\ref{tab:TF}.
}
\end{figure}

\begin{figure}[t]

\vspace{8mm}
\includegraphics[width=0.5\textwidth]{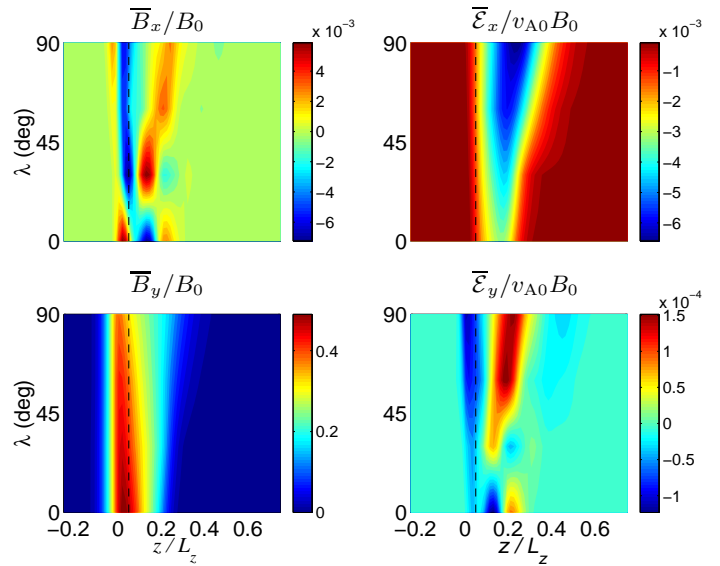}\\[-7.5cm]
\hspace*{0.18\columnwidth}$\meanB_{x}/B_0$\\[3.1cm]
\hspace*{0.18\columnwidth}$\meanB_{y}/B_0$\\[-3.9cm]
\hspace*{0.68\columnwidth}$\meanemf_{x}/\vAN B_0$\\[3.1cm]
\hspace*{0.68\columnwidth}$\meanemf_{y}/\vAN B_0$\\[3.5cm]
\caption{\label{fig:emf_thetadep0} Latitudinal dependence of
$\meanB_{x,y}(z)$, $\meanemf_{x,y}(z)$
averaged between $t=t^{\rm sat}$ and $t^{\rm sat}+t_{\rm A0}$.
All problem parameters except $\theta$ held fixed at the values
of run TF$0+$ in Table~\ref{tab:TF}.
Dashed line: initial position of the magnetic layer.}
\end{figure}

\begin{figure}

\vspace{1cm}
\includegraphics[width=0.5\textwidth]{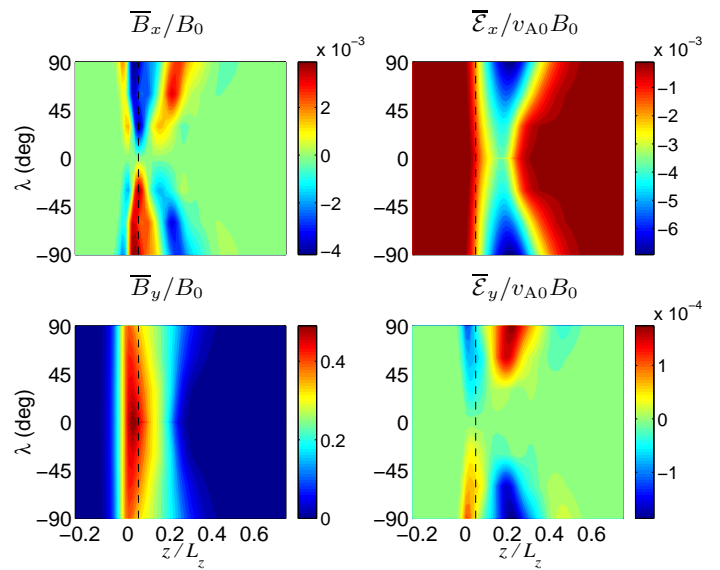}\\[-7.5cm]
\hspace*{0.18\columnwidth}$\meanB_{x}/B_0$\\[3.1cm]
\hspace*{0.18\columnwidth}$\meanB_{y}/B_0$\\[-3.9cm]
\hspace*{0.68\columnwidth}$\meanemf_{x}/\vAN B_0$\\[3.1cm]
\hspace*{0.68\columnwidth}$\meanemf_{y}/\vAN B_0$\\[3.5cm]
\caption{\label{fig:emf_thetadep} Same as Fig.~\ref{fig:emf_thetadep0} but now symmetrized about the equator.}
\end{figure}

\begin{figure}[t!]
\includegraphics[width=0.5\textwidth]{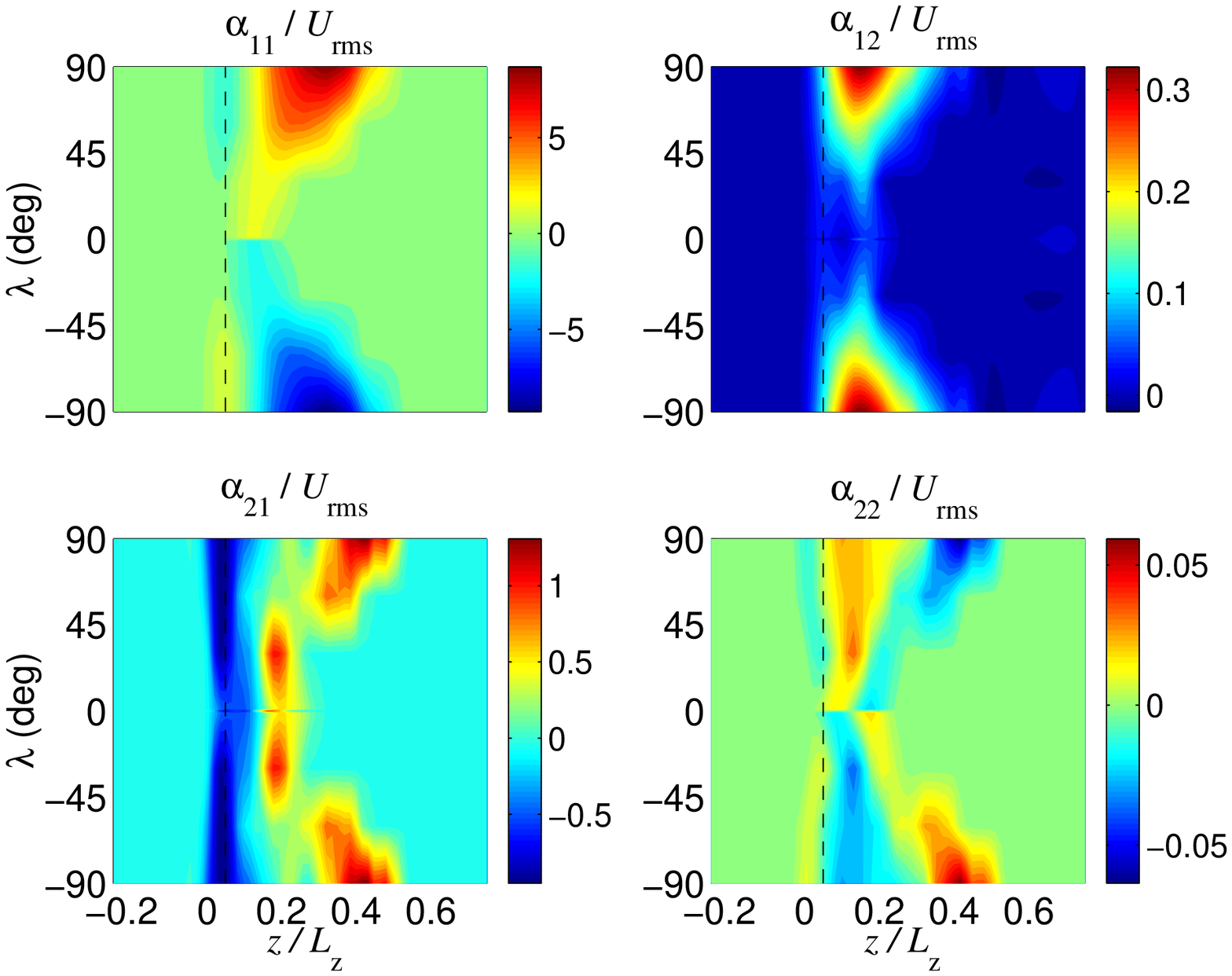}
\includegraphics[width=0.5\textwidth]{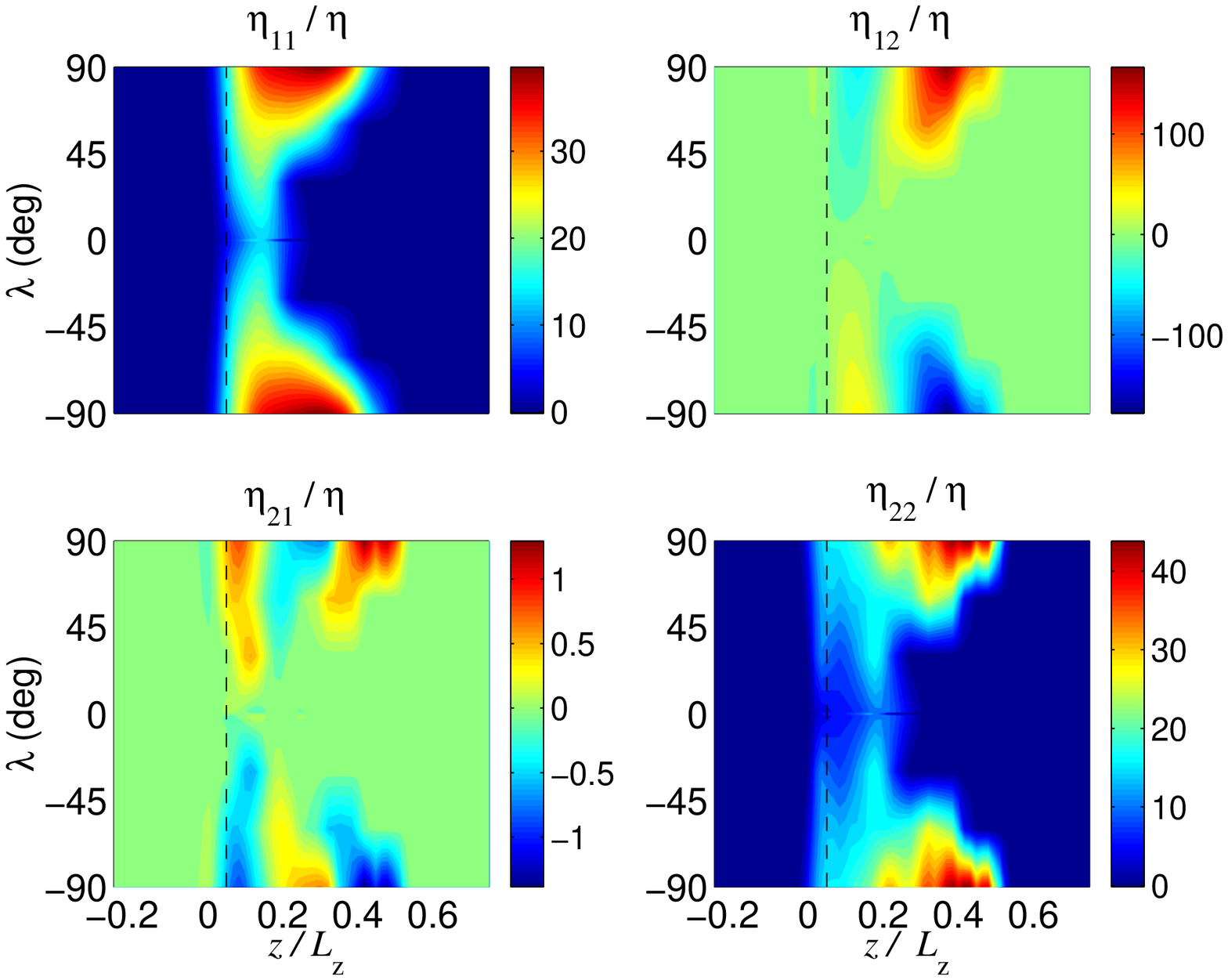}
\caption{\label{fig:alp_thetadep} Dependence of
$\aTens$ and $\eTens$,
averaged between $t=t^{\rm sat}$ and $t^{\rm sat}+t^{\rm A0}$,
on latitude, $\lambda = 90^\circ-\theta$, and $z$, 
calculated using test fields with
$k' = 0.5$.
All problem parameters except $\theta$ held fixed at the values of run TF$0+$ in Table~2.
$\aTens$ scaled by $U_\rms$, 
$\eTens$  scaled by the molecular
diffusivity
$\eta$.
Dashed line:  initial
position of the magnetic layer.}
\end{figure}

According to the results of \cite{Sch00} we expect 
a decrease in the intensity of the instability with increasing inclination of the rotation axis.
This can be explained by the buoyant nature of the turbulence, for which vertical motions are essential.
At the poles, the effect of the Coriolis force on vertical motions is weakest,
whereas they are strongly deflected at the equator.
\Fig{fig:thetadep} indeed confirms, that the growth rates  decrease continuously when changing 
$\theta$ from $0^\circ$ towards $90^\circ$.
In Fig.~\ref{fig:emf_thetadep0} we show the 
variation of the
mean magnetic field and the corresponding mean EMF with latitude and $z$
at a time during the saturated stage. 
A pecularity in this figure is
that $\meanemfs_y$ and consequently $\meanB_x$ are non-zero
at the equator where we would expect 
these quantities to vanish.
This is an example of spontaneous symmetry breaking and can be explained by a mean field dynamo
operating at the equator. This dynamo generates $\meanB_x$ whose sign is determined by the random 
initial conditions. A detailed discussion of this issue will be provided in a forthcoming paper.
In the rest of the paper we anti-symmetrize $\meanB_x$ and $\meanemfs_y$,
while symmetrize $\meanB_y$ and $\meanemfs_x$ about the equator (see Fig.~\ref{fig:emf_thetadep}). This is done by including the results
from runs with two different initial conditions for velocity, one being the mirror reflection of the other
according to Eq.~(\ref{eq:refl}). In particular at the equator ($\theta=90^\circ$), 
the two initial conditions give 
rise to a $\meanB_x$ with exactly the same magnitude but differing in sign. 
Thus averaging the $\meanB_x$ from the two runs
gives a zero $\meanB_x$ at the equator.
We perform the same operation for the turbulent 
transport coefficients calculated from the QKTF method.
The transport coefficients
calculated from only the $k'=0.5$ test fields 
belonging to the 
family of cosine modes are presented in 
Fig.~\ref{fig:alp_thetadep}.
It can be immediately seen 
from these plots that the instability becomes more effective with 
increasing
(northern or southern) latitude.
Corresponding runs performed with linear test-fields
are compiled in Table~\ref{tab:TF}.
We observe that the turbulent transport coefficients increase in modulus
when moving towards the poles, but are, with the only exception of
$\alpha_{21}$, significantly reduced close to the equator.
Obviously,
the transport coefficients
respond
directly
to the
inhibition
of the vertical motions 
by the Coriolis force when moving towards the equator.
\begin{table}[t]
\caption{\label{tab:TF} List of runs from set TF. 
$\Pr = 4.0$, $\Pm=4.0$, $\Tm=3.24\times10^{10}$, except for last three runs with $\Tm=0$.
$\betaZ=2.27$ except in
last two runs,
TF00l and TF00m, where $\betaZ$ is 3.22 and 1.03, respectively.
Resolution $64^3$ throughout.
Saturation is reached at $t^{\rm sat}$.
Global extrema of the 
dominating $\meanemfs_y$
with respect to $z$ and $t$ are given.
$\Rey \sim 4.8$ throughout. 
The arrows in the $\Omega$ column indicate the sign of $\Omega_z$.
}
\centering
\begin{tabular}{@{\hspace{0mm}}c@{\hspace{0mm}} l  c c c @{\hspace{5.5mm}} c@{\hspace{-0.5mm}} c@{}}     
\hline\hline\\[-2mm]
&\phantom{a}Set  & $\theta/^\circ$  & $\Omega$ &  $t^{\rm sat}/t_{\mathrm{A0}}$ & 
\multicolumn{2}{c}{\hspace*{-5mm}$10^4 \times \meanemfs_y /\vAN B_{y0}$}\\   
 &&&&& min & max \\[1mm] 
\hline\\[-2mm]
  & TF0$+$  & 0 & $\uparrow$ &  2.42 & $-1.82$&3.52  \\   
  & TF0$-$ & 0 & $\downarrow$ &  2.42 & $-4.41$ & 1.44 \\  
  & TF30$+$ & 30 & $\uparrow$ &  2.58&  $-1.71$ & 3.68 \\
  & TF60$+$ & 60 & $\uparrow$ &  2.87 & $-1.33$ & 3.16\\
  & TF89$+$ & 89 & $\uparrow$ &  3.00 &  $-1.21$ & 2.10\\
  & TF90$+$ & 90 & $\uparrow$ &  3.50 &  $-1.35$ & 1.62\\
  & TF00 & 0 & 0 &  2.21 & $-3.25$ & 3.12\\
  & TF00l & 0 & 0 &  3.12 & $-0.84$ & 1.38\\
  & TF00m & 0 & 0 &  2.20 & $-4.53$ & 8.43\\[1mm]
\hline
\end{tabular}
\end{table}

\begin{figure}
\includegraphics[width=\columnwidth]{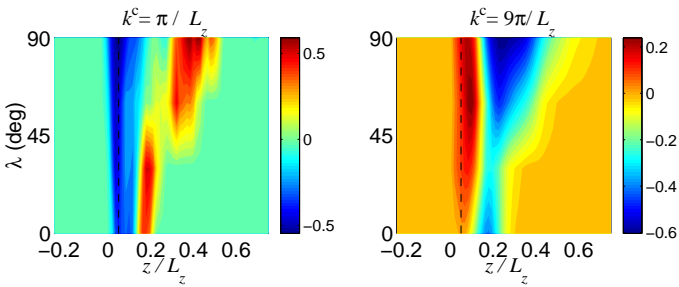}
\caption{\label{fig:gamma} Pumping velocity
$\gamma_z = (\alpha_{21} - \alpha_{12})/2$
 scaled with $U_\rms$, 
as a function of latitude, $\lambda$ for $k'=0.5$ (left panel) 
and $k'= 4.5$ (right panel).}
\end{figure}

The $\alpha$ tensor can be decomposed in symmetric and antisymmetric parts. The latter
represents a {\em turbulent pumping velocity} $\boldsymbol{\gamma}$, and 
gives rise to the term $\boldsymbol{\gamma} \times \meanBB$ in the mean EMF. By virtue of 
the horizontal averaging of the magnetic field, $\meanB_z=0$. Hence, the only relevant 
component of pumping is $\gamma_z$ which
is defined by $(\alpha_{21}-\alpha_{12})/2$. 
Analytical results indicate that 
in a wide range of situations,
the turbulent pumping
is directed away from the region of strong turbulence
\citep[``turbulent diamagnetism'', see][]{KR80}. From Fig.~\ref{fig:kdep}, we see that 
the components $\alpha_{21}$ and $\alpha_{12}$ do not converge to zero
with increasing $k'$. 
In fact $\alpha_{12}$ changes sign at $k'=2$ and $\alpha_{21}$ does so at $k'=8$.
Consequently ${\gamma_z}$ determined 
from  harmonic test fields with $k'=0.5$ and $k' \geq 4$ 
should have opposite signs as
confirmed by
Fig.~\ref{fig:gamma}.
Physically, this means that magnetic fields formed on the scale of $L_z$ 
will be pumped away from the initial magnetic layer while those on the scale of the 
magnetic layer, $H_B$
shall be pumped into the layer, the latter being contrary to the
standard concept
of ``turbulent diamagnetism".
It is thus difficult to comment on the transport of the
total
$\meanBB$ by 
$\boldsymbol{\gamma}$.
Only if the pumping
were
oriented away from the magnetic layer for
all the wavenumbers of the dominating constituents in $\meanBB$
it would
lead to a broadening of the initial layer i.e., a reduction of 
$\partial B_y/\partial z$
and would hence inhibit the instability.
A similar dependence of turbulent pumping on wavenumber has been found
by \cite{kap+kor+bra09} in DNS of convection. 
With regard to
to the saturation of the magnetic buoyancy instability,
a strong turbulent magnetic 
diffusion given by $\eta_{22}$ (see Fig.~\ref{fig:thetadep}) 
is likely to be more important.
At the poles this quantity 
is as large as $40$ times the molecular value of $\eta$.  

\section{Conclusions}

We have studied in detail the generation of the $\alpha$ effect due to
the
buoyancy
instability of a toroidal  magnetic layer in a stratified atmosphere by using direct numerical
simulations. We find that both the 
magnetic energy and the current helicity in the system 
increase monotonically with the ratio of thermal conductivity
to magnetic diffusivity,
the Roberts number
$\Rb$ (Fig.~\ref{fig:prm}).
This agrees with earlier analytical
work of  \cite{Gil70} and \cite{Ach79} as well as numerical work of 
\cite{Silvers} which find that efficient thermal diffusion 
or heat exchange can destabilize a stable stratification.
The dependence of twist on $\Rb$
is an important result since the buoyancy instability
would produce
twisted flux tubes from a magnetic layer, if
it existed
in the overshoot layer of the Sun. 
\cite{VB08} also reported the formation of twisted flux tubes from a horizontal
magnetic layer produced, but in their case it is
due to the action of shear on a weak vertical magnetic field.
We further find that the growth rate of the buoyancy instability is reduced in
presence of rotation compared to the case with $\Omega=0$. 
 
We have run our simulations only until the time taken by the initial
magnetic layer to break up due to the buoyancy instability.
In absence of any other forcing such as a strong shear, the buoyancy
instability cannot usually sustain itself past the break-up phase
since the vertical gradient of the magnetic energy in the layer becomes
comparable to the stratification due to magnetic diffusion.
We may say that strong shear is not imperative to the production of
tubular structures from the toroidal magnetic layer but will play a key
role in keeping the layer from breaking up.
It may also be possible that turbulent pumping arrests the decay of
such a magnetic layer in the actual overshoot region.
However, it is not yet clear if such a layer
exists and is subject
to the buoyancy instability in the real Sun.

We have `measured' the turbulent transport coefficients using the
technique of the quasi-kinematic test-field method. In order to prove that the $\aTens$ and $\eTens$ 
tensors
obtained from this method are reasonably accurate, we show the agreement 
between $\meanEMF=\overline{\vec{u'}\times\vec{b'}}$ and the ansatz 
$\meanEMF=\aTens\overline{\vec{B}}-\eTens\overline{\vec{J}}$ using 
harmonic 
test fields with
wavenumbers
$0\le k'\le 16$. 
Here we have 
illustrated a technique of judging the reliability of transport coefficients obtained
from the test-field method. 
We find that, even in presence of magnetically driven turbulence,
$\aTens$ and $\eTens$ obtained from the quasi-kinematic test-field method
provide a reasonably accurate description of the turbulent EMF.
This is an important outcome of our study.

We find that $\meanemfs_x$ determined using a
harmonic test field with the lowest wavenumber that fits in vertical extent of the box 
already comprises a considerable part of the total EMF. 
Hence we can use QKTF 
to calculate the turbulent coefficients at finite $\Omega$ as a function of latitude 
using harmonic test fields with this wavenumber. 
The component $\alpha_{22}$ contributes to the generation of
$\overline{B}_x$ from the strong initial field $\overline{B}_y$ in the layer.
The off-diagonal components contribute to a vertical turbulent 
pumping velocity directed away from the region of turbulence surrounding
the magnetic layer.
The influence of this component 
systematically expands along $z$ with increasing latitude and somewhat agrees
 with the result in Brandenburg \& Schmitt (1998). 
The agreement is not complete since the $\alpha_{22}(z, \theta)$
is inhomogeneous with respect to $z$ and can have sign changes along
$\theta$, e.g., at $z/L_z=0.4$ in Fig.~\ref{fig:alp_thetadep}d.
We find that all transport coefficients except $\alpha_{21}$ increase
with latitude and are significantly reduced near the equator due to the
suppressing effect of the Coriolis force on the instability.

For the first time the turbulent magnetic diffusivity given by the diagonal
components of $\eTens$ has been computed, as shown in Fig.~\ref{fig:alp_thetadep}.
In particular, near the magnetic layer, the diagonal component 
$\eta_{22}$ is 25 times larger than the molecular value $\eta$. 
The buoyancy driven instability has the property that the $\alpha$ as measured 
by the growth rate of the instability 
increases with the magnitude of the magnetic field in the horizontal
layer (compare solid and dashed lines Fig.~\ref{fig:omgdep}). 
This property makes it an attractive candidate for solar dynamo models, unlike
the $\alpha$ generated due to helical turbulence which gets quenched
for strong magnetic fields. 
The increase of $\alpha$ and $\eta$ with $\meanBB$ is a remarkable result 
and supports similar suggestions by \cite{BST98} that, if turbulent transport
coefficients are caused by flows that are magnetically driven like here or,
e.g., in Balbus-Hawley instabilities, then both $\alpha$ and $\eta$ may
increase with the magnetic field strength.
This trend is sometimes referred to as `anti-quenching' and may be needed
to support the observational relation between the ratio of dynamo cycle
to rotation frequencies, $\omega_{\rm cyc}/\Omega$ and
Rossby number inverse, $\rm{Ro}^{-1}$ for stellar data 
\citep{BST98,SB99}.
Note finally that  modelling the $\alpha$ as a function of space and 
the mean magnetic field to use in a mean field dynamo model 
is a very difficult proposition that needs to be postponed to future work.

\begin{acknowledgements}
We thank A. Hubbard for reading the manuscript carefully. 
The computations have been carried out on the
National Supercomputer Centre in Link\"oping and the Center for
Parallel Computers at the Royal Institute of Technology in Sweden.
This work was supported in part by
the European Research Council under the AstroDyn Research Project No.\ 227952
and the Swedish Research Council Grant No.\ 621-2007-4064.
\end{acknowledgements}


\begin{thebibliography}{}

\bibitem[{Acheson(1979)}]{Ach79}
Acheson, D. J. 1979, Sol. Phys. 62,23

\bibitem[{Brandenburg(1998)}]{}
Brandenburg, A.\yproc{1998}{61}
{Theory of Black Hole Accretion Discs}
{M. A. Abramowicz, G. Bj\"ornsson \& J. E. Pringle}
{Cambridge University Press}

\bibitem[{Brandenburg et al.(1998)}]{BST98}
Brandenburg, A., Saar, S. H., \& Turpin, C. R.\yapj{1998}{498}{L51}

\bibitem[{Brandenburg \& Schmitt(1998)}]{BS98}
Brandenburg, A., \& Schmitt, D.\yana{1998}{338}{L55}

\bibitem[{Brandenburg et al.(1995)}]{BNST}
Brandenburg, A., Nordlund, \AA., Stein, R. F., \& Torkelsson, U.\yapj{1995}{446}{741}

\bibitem[{Brandenburg et al.(2008a)}]{BRS08}
Brandenburg, A., R\"adler, K.-H., \& Schrinner, M.\yana{2008a}{482}{739}

\bibitem[{Brandenburg et al.(2008b)}]{BRRK08}
Brandenburg, A., R\"adler, K.-H., Rheinhardt, M., \& K\"apyl\"a, P. J.\yapj{2008b}{676}{740}

\bibitem[{Brandenburg et al.(2008c)}]{BRRS08}
Brandenburg, A., R\"adler, K.-H., Rheinhardt, M., \& Subramanian, K.\yapjl{2008c}{687}{L49}

\bibitem[{Brandenburg et al.(2010)}]{Betal10}
Brandenburg, A., Chatterjee, P., Del Sordo, F., Hubbard, A., K\"apyl\"a, P. J., \& Rheinhardt, M.\pjour{2010}{Physica Scripta T}

\bibitem[{Cline et al.(2003)}]{CBC03}
Cline, K. S., Brummell, N. H., \& Cattaneo, F.\yapj{2003}{599}{1449}

\bibitem[{Courvoisier et al.(2010)}]{Courvoisier10}
Courvoisier A., Hughes D. W., Proctor M. R. E.\yprs{2010}{466}{583}

\bibitem[{Fan(2001)}]{Fan01}
Fan, Y.\yapj{2001}{546}{509}

\bibitem[{Gilman(1970)}]{Gil70}
Gilman, P. A., \yapj{1970}{162}{1019}

\bibitem[{Hubbard \& Brandenburg(2009)}]{hub+bra09}
Hubbard, A., \& Brandenburg, A.\yapj{2009}{706}{712}

\bibitem[{K\"apyl\"a, Korpi \& Brandenburg(2009)}]{kap+kor+bra09}
K\"apyl\"a, P.J., Korpi, M., \& Brandenburg, A.
\yana{2009}{500}{633}


\bibitem[{Krause \& R\"adler(1980)}]{KR80}
Krause F., \& R\"adler K.-H., 1980, 
Mean-Field Magnetohydrodynamics and Dynamo Theory (Pergamon Press, Oxford)

\bibitem[{Matthews et al.(1995)}]{MHP95}
Matthews, P. C., Hughes, D. W., \& Proctor, M. R. E.\yapj{1995}{448}{938}

\bibitem[{Moffatt(1978)}]{Mof78}
Moffatt, H. K.\yjfm{1972}{53}{385}

\bibitem[{Rheinhardt \& Brandenburg(2010)}]{RB10}
Rheinhardt, M., \& Brandenburg, A.\yana{2010}{520}{A28}

\bibitem[{Saar \& Brandenburg(1999)}]{SB99}
Saar, S. H., \& Brandenburg, A.\yapj{1999}{524}{295}

\bibitem[{Schmitt(1984)}]{Sch84}
Schmitt, D.\yproc{1984}{223}
{ESA}{The Hydromagnetics of the Sun}{N85-25091 14-92}

\bibitem[{Schmitt(1985)}]{Sch85}
Schmitt, D.\ybook{1985}{Dynamowirkung magnetostrophischer Wellen}
{PhD thesis, University of G\"ottingen}

\bibitem[{Schmitt(2000)}]{Sch00}
Schmitt, D.\yproc{2000}{}
{The fluid mechanics of astrophysics and geophysics}
{Advances in Nonlinear dynamos}
{A. Ferris-Mas, M. N\'u\~nez Jin\'enez}

\bibitem[{Schrinner {et al.}(2005)}]{Sch05}
Schrinner, M., R\"adler, K.-H., Schmitt, D., Rheinhardt, M.,
Christensen, U. 2005, Astron. Nachr., 326, 245

\bibitem[{Schrinner {et al.}(2007)}]{Sch07}
Schrinner, M., R\"adler, K.-H., Schmitt, D., Rheinhardt, M.,
Christensen, U. 2007, GAFD, 101, 81

\bibitem[{Silvers et al.(2009)}]{Silvers}
Silvers, L. J., Vasil, G. M., Brummel, N. H., \& Proctor, M. R. E.\yapj{2009}{702L}{14}

\bibitem[{Vasil \& Brummel(2008)}]{VB08}
Vasil, G. M., \& Brummel, N. H. \yapj{2008}{686}{709}

\bibitem[{Vermersch \& Brandenburg(2009)}]{VB09}
Vermersch, V., \& Brandenburg, A.\yan{2009}{330}{797}

\end{thebibliography}
\end{document}